\documentclass[12pt,dvipdfm,twoside]{article}
\usepackage[dvips]{graphicx}
\usepackage{amsthm, amssymb, epsfig}
\usepackage[usenames,dvipsnames]{color}
\usepackage[english]{babel}
\usepackage{fancyhdr}
\usepackage{verbatim}

\def\myPrimaryAuthor{Andrey Itkin}
\def\myAuthor{Andrey Itkin
}
\def\myProjectName{Pricing options with VG model using FFT}
\def\myProjectNum{}
\def\myAbsract{
We discuss various analytic and numerical methods that have been used to get option prices within a framework of the VG model. We show that some popular methods, for instance, Carr-Madan's FFT method \cite{CarrMadan:99a} could blow up for certain values of the model parameters even for an European vanilla option. Alternative methods - one originally proposed by Lewis, and Black-Scholes-wise method are considered that seem to work fine for any value of the VG parameters. Test examples are given to demonstrate efficiency of these methods. Convergency of all methods is also discussed.}

\def\ni{\noindent}
\def\E{{\mathbb E}}
\def\bi{\begin{itemize}}     \def\ei{\end{itemize}}
\def\be{\begin{enumerate}}   \def\ee{\end{enumerate}}
\def\beq{\begin{equation}}   \def\eeq{\end{equation}}
\def\bd{\begin{displaymath}} \def\ed{\end{displaymath}}
\def\bea{\begin{eqnarray}}   \def\beaz{\begin{eqnarray*}}
\def\eea{\end{eqnarray}}     \def\eeaz{\end{eqnarray*}}
\def\beac{\begin{eqnarrayc}} \def\eeac{\end{eqnarrayc}}
\def\nn{\nonumber}           
\def\dsize{\displaystyle}    \def\dfrac#1#2{\frac{\dsize #1}{\dsize #2}}
\def\sfrac#1#2{\frac{#1}{#2}}

{\catcode`$=9 $ 
}

   {\noindent \footnotesize \verbatim}%
   {\normalsize \endverbatim}

\normalsize

\begin{document}

\pagestyle{plain}
\vspace{-2cm}
\begin{figure}
\begin{minipage}[t]{\textwidth}
\DeclareGraphicsRule{.jpg}{bmp}{.bb}{}
\hspace{-3mm}
\end{minipage}
\end{figure}

\title{\vspace{2cm}\color{blue}\myProjectName \break \break \myProjectNum}
\author{\color{red}\myAuthor}
\date{}
\maketitle

\begin{center}
{\color{blue} Rutgers University, New Jersey, \\
Department of mathematics\\
aitkin@math.rutgers.edu}
\end{center}
\smallskip

\renewcommand\abstractname{Summary}
\abstract{\myAbsract}

\newpage
\pagestyle{fancyplain}
\fancyfoot[RE,RO]{\thepage}
\fancyfoot[LE,LO]{\color{red}\myPrimaryAuthor\hfill}
\fancyfoot[CE,CO]{\color{red} \footnotesize{\myProjectName \\
\myProjectNum}}

\def\itemDesc#1#2#3{
    \colorbox{#1} {\parbox{0.9\columnwidth}{\color{#2} {\scriptsize #3}}\\}
}
\setcounter{tocdepth}{7}
\tableofcontents
\newpage

\section{Introduction}
This paper summarizes some results of work originally initiated by
Peter Carr. It supposes to investigate various numerical and
analytical methods of option pricing using VG model in order to find
out which algorithm is most efficient.

Let us first give a brief overview of the VG model. The Variance Gamma
(VG for short) process was proposed by Madan and Seneta (see
\cite{MadanSeneta1990}) to describe stock price dynamics instead of the
Brownian motion in the original Black-Scholes model. Two new
parameters: $\theta$ skewness and $\nu$ kurtosis are introduced in
order to describe asymmetry and fat tails of real life distributions.
The VG process is defined by evaluating Brownian motion with drift at a
random time specified by gamma process. In other words, the VG model
with parameter vector $(\sigma, \nu, \theta)$ assumes that the forward
price satisfies the following equation

\beq \label{underVG} \ln F_{t} = \ln F_{0} + X_{t} + \omega t, \eeq
where \beq X_{t} = \theta \gamma_{t}(1, \nu) + \sigma W_{\gamma_{t}(1,
\nu)}, \eeq

\ni and $\gamma_{t}(1, \nu)$ is a Gamma process playing the role of
time in this case with unit mean rate and density function given by

\beq \label{VGdensity} f_{\gamma_{t}(1, \nu)}(x) =
\frac{x^{\frac{t}{\nu}-1}e^{-\frac{x}{\nu}}}
{\nu^{\frac{t}{\nu}}\Gamma\left(\frac{t}{\nu}\right)}. \eeq

In the Eq.~(\ref{underVG}) $\omega$ is chosen to make $F_{t}$ a
martingale.

The probability density function for the VG process may be written as

\begin{equation}\label{pdfVG}
  h_t(x) = \int_0^\infty \dfrac{dg}{\sqrt{2\pi g}}\exp \left[
  - \dfrac{(x-\theta g)^2}{2 \sigma ^2g}\right]
  \frac{g^{\frac{t}{\nu}-1}e^{-\frac{g}{\nu}}}
{\nu^{\frac{t}{\nu}}\Gamma\left(\sfrac{t}{\nu}\right)}
\end{equation}

\ni or after integration over $g$

\begin{equation}\label{pdfVGfin}
  h_t(x) = \dfrac{2\exp \left( \theta x /\sigma ^2 \right)}{\sqrt{2}\pi \sigma
  \nu^{\frac{t}{\nu}} \Gamma\left(\sfrac{t}{\nu}\right)} \left( \dfrac{x^2}{\theta ^2 +
\sfrac{2\sigma ^2}{\nu}} \right)^{\frac{t}{2\nu} - \frac{1}{4}}
K_{\frac{t}{\nu} - \frac{1}{2}}\left( \dfrac{1}{\sigma ^2}
\sqrt{x^2\left( \theta ^2 + \frac{2\sigma ^2}{\nu}\right)} \right),
\end{equation}

\ni where $K$ is the modified Bessel function of the second kind. The
characteristic function $ \phi_{\gamma_{t}(1, \nu)}(u)$ for the VG
process has remarkably simple form

\begin{equation}\label{char1}
  \phi _t(u) \equiv \left< E^{iux} \right> \equiv \int_0^\infty
  h_t(x)e^{iux} dx = \dfrac{1} {(1 - i\theta \nu u + \frac{1}{2}\sigma ^2\nu
  u^2)^\frac{t}{\nu}}.
\end{equation}

Another derivation of this expression could be obtained when
conditioning on time change like in Romano-Touzi for stochastic
volatility models

\bea \label{charFunc}
\phi_{\gamma_{t}(1, \nu)}(u) &=&
    \E[e^{iu\gamma_{t}(1, \nu)}] = \int_{0}^{\infty}e^{iux}f_{\gamma_{t}(1, \nu)}(x)dx
    = \int_{0}^{\infty}e^{iux}\frac{x^{\frac{t}{\nu}-1}e^{-\frac{x}{\nu}}}{\nu^{\frac{t}{\nu}}
    \Gamma\left(\frac{t}{\nu}\right)}dx
\nn\\ &=&
    \int_{0}^{\infty}\frac{x^{\frac{t}{\nu}-1}e^{-\frac{x(1 - iu\nu)}{\nu}}}{\nu^{\frac{t}{\nu}}
    \Gamma\left(\frac{t}{\nu}\right)}dx
\nn\\ &=&
    (1 - iu\nu)^{-\frac{t}{\nu}}\int_{0}^{\infty}\frac{(x(1 - iu\nu))^{\frac{t}{\nu}-1}
    e^{-\frac{x(1 - iu\nu)}{\nu}}}{\nu^{\frac{t}{\nu}}\Gamma\left(\frac{t}{\nu}\right)}
    d(x(1 - iu\nu))
\nn\\ &=&
    (1 - iu\nu)^{-\frac{t}{\nu}}\int_{0}^{\infty}\frac{y^{\frac{t}{\nu}-1}
    e^{-\frac{y}{\nu}}}{\nu^{\frac{t}{\nu}}\Gamma\left(\frac{t}{\nu}\right)}dy
    = (1 - iu\nu)^{-\frac{t}{\nu}}.
\eea

\bea \label{interm1}
    \phi_{X_{t}}(u)   &=& \E[e^{iuX_{t}}] =
    \E[\E[e^{iuX_{t}} \mid \gamma_{t}(1, \nu)]] =
    \E[\E[e^{iu\left(\theta \gamma_{t}(1, \nu) + \sigma W_{\gamma_{t}(1, \nu)}\right)}
    \mid \gamma_{t}(1, \nu)]]
\nn\\ &=&
    \E[e^{iu\theta \gamma_{t}(1, \nu) - \frac{1}{2}u^{2}\sigma^{2}\gamma_{t}(1,
    \nu)}] = \E[e^{i\left(u\theta + i\frac{1}{2}u^{2}\sigma^{2}\right)\gamma_{t}(1,
    \nu)}] \nn\\&=& \phi_{\gamma_{t}(1, \nu)}(u\theta  + i\frac{1}{2}u^{2}\sigma^{2})
    = \left( 1 - i\theta\nu u + \frac{1}{2}\sigma^{2}\nu u^{2}\right)^{-\frac{t}{\nu}}.
\eea

Now, to prevent arbitrage, we need $F_{t}$ be a martingale, and, since
$F_{t}$ is already an independent increment process, all we need is

\beq \E[F_{t}] = F_{0}, \eeq

\ni or

\beq \E[F_{0}e^{X_{t} + \omega t}] = F_{0}\phi_{X_{t}}(-i)e^{\omega t}
= F_{0}. \eeq

This tells us that

\beq \label{omega} \omega = - \frac{\ln \phi_{X_{t}}(-i)}{t} =
-\frac{-\frac{t}{\nu} \ln\left( 1 - \theta\nu -
\frac{1}{2}\sigma^{2}\nu \right)}{t} =\frac{1}{\nu}\ln\left( 1 -
\theta\nu - \frac{1}{2}\sigma^{2}\nu \right). \eeq

Note that from the definition of $\omega$ above, in order to have  a
risk neutral measure for VG model, its parameters must obey an
inequality:

\beq \label{constrain} \dfrac{1}{\nu} > \theta + \frac{\sigma^{2}}{2}.
\eeq

Note that risk neutral parameters $\theta , \nu, \sigma $ do not have
to be equal to their statistical counterparts.

Accordingly, the characteristic function of the $x_T \equiv \log
S_T$ VG process is

\beq \label{charS}
    \phi(u)   = \dfrac{S_0 e^{(r-q+\omega)T}}{ \left( 1 - i\theta\nu u + \frac{1}{2}\sigma^{2}\nu u^{2}\right)^{\frac{T}{\nu}}}.
\eeq

Statistical parameters of VG distribution may be calculated from the
historical data on stock prices. In particular we have to find the
values of the parameters $\theta ^*, \nu^*$ and $\sigma ^*$ such that
the folloiwng expression is maximized:

\begin{equation}\label{calibr}
  \prod_{j=1}^{n} h_{\tau _j} (x_j),
\end{equation}

\ni where $h_{\tau _j} (x_j)$ are given by Eq.\ref{pdfVGfin} and $x_j$
are observed returns per time $\tau _j$, i.e. $x_j =
\log(S_j/S_{j-1})$.

\section{Pricing European option}

The value of European option on a stock when the risk neutral
dynamics is given by Eq.~(\ref{underVG}) is

\begin{equation}\label{EurVG}
  V = \exp(-rT) \int_{-\infty}^{\infty} h_T\left(x-(r-q+\omega )T\right)
  W(e^x)dx,
\end{equation}

\ni where $T$ is time until expiration, $q$ is continuous dividend and
$W(e^x)$ is payoff function that has the following form

\begin{equation}\label{payoff}
  W(e^x) = (S_0e^x - K)^+ - \mbox{call}, \quad W(e^x) = (K - S_0e^x)^+ -
  \mbox{put}.
\end{equation}

Direct calculation allows us to derive the put-call parity relation
identical to Black-Scholes case

\begin{equation}\label{parity}
  C = S_0 e^{-qT} - Ke^{-rT} + P.
\end{equation}

There are several methods to price a European option under the VG
model. One method uses the closed form solution derived in
\cite{MadCarrChang}. Although the expression is analytic it requires
computation of modified Bessel functions, and hence may not be as fast
as we would like our pricing model to be. Therefore, FFT method has
been widely utilized to obtain a more efficient pricer. Few flavors of
the FFT method has been previously discussed with regard to the VG model.

First of all the FFT method of Carr and Madan \cite{CarrMadan:99a},
nowadays almost standard in math finance, was applied to the VG model
to price the European vanilla option since the characteristic function
of the log-return process has a very simple form given above. Further
we intend to show, that unfortunately this method blows up at some
values of the VG parameters.

Mike Konikov and Dilip Madan \cite{MadanKonikov:2004a} proposed another
interesting method based on the definition of the VG process as being a
time changed Brownian motion, where the time change is assumed
independent of the Brownian motion. This method was described in detail
in \cite{MadanKonikov:2004a} while has not been implemented yet.


Also Mike Konikov and I independently implemented a modification of the FFT
method - the Fractional Fourier Transform, which is described in detail
in \cite{Bailey_Swarztrauber_1991, Chourdakis2004}. This method usually allows
acceleration of the pricing function by factor 8-10, while for the VG model it
still demonstrates same problem as the original FFT.

Below we discuss why the Carr and Madan FFT approach fails for the VG model.
We propose another method, which originally has been developed in a general form by Lewis \cite{Lewis:2001},
that seems to be free of such problems.

\section{Carr-Madan's FFT approach and the VG model}

Let us start with a short description of the Carr-Madan FFT method. It was
worked out for models where the characteristic function of underlying price
process ($S_t$) is available. Therefore, the vanilla options can be priced very
efficiently using FFT as described in Carr and Madan ~\cite{CarrMadan:99a}. The
characteristic function of the price process is given by

\begin{equation}\label{cFunc}
\phi(u,t)=\E(e^{iuX_t}),
\end{equation}

\ni  where $X_t=\log(S_t)$. Note that the above representation holds for all
models and is not just restricted to L\'evy models where the characteristic
functions have a time homogeneity constraint that $\phi(u,t)=e^{-t\psi(u)}$,
where $\psi(u)$ is the L\'evy characteristic exponent.

Once the characteristic function is available, then the vanilla call option can
be priced using Carr-Madan's FFT formula:

\begin{equation} \label{callFFT}
C(K,T)=\frac{e^{-\alpha\log(K)}}{\pi} \int_0^{\infty}\mathrm{Re}
\left[e^{-iv\log(K)}\omega(v)\right]dv,
\end{equation}

\ni where

\begin{equation} \label{omega}
\omega(v)=\frac{e^{-rT}\phi(v-(\alpha+1)i,
T)}{\alpha^2+\alpha-v^2+i(2\alpha+1)v}
\end{equation}

The integral in the first equation can be computed using FFT, and as a result
we get call option prices for a variety of strikes. For complete details, see
Carr \& Madan paper \cite{CarrMadan:99a}.

The put option values can just be constructed from Put-Call
symmetry.
\begin{figure}[ht]
\begin{flushleft}
\begin{minipage}[ht]{0.4\textwidth}
\includegraphics[totalheight=2in]{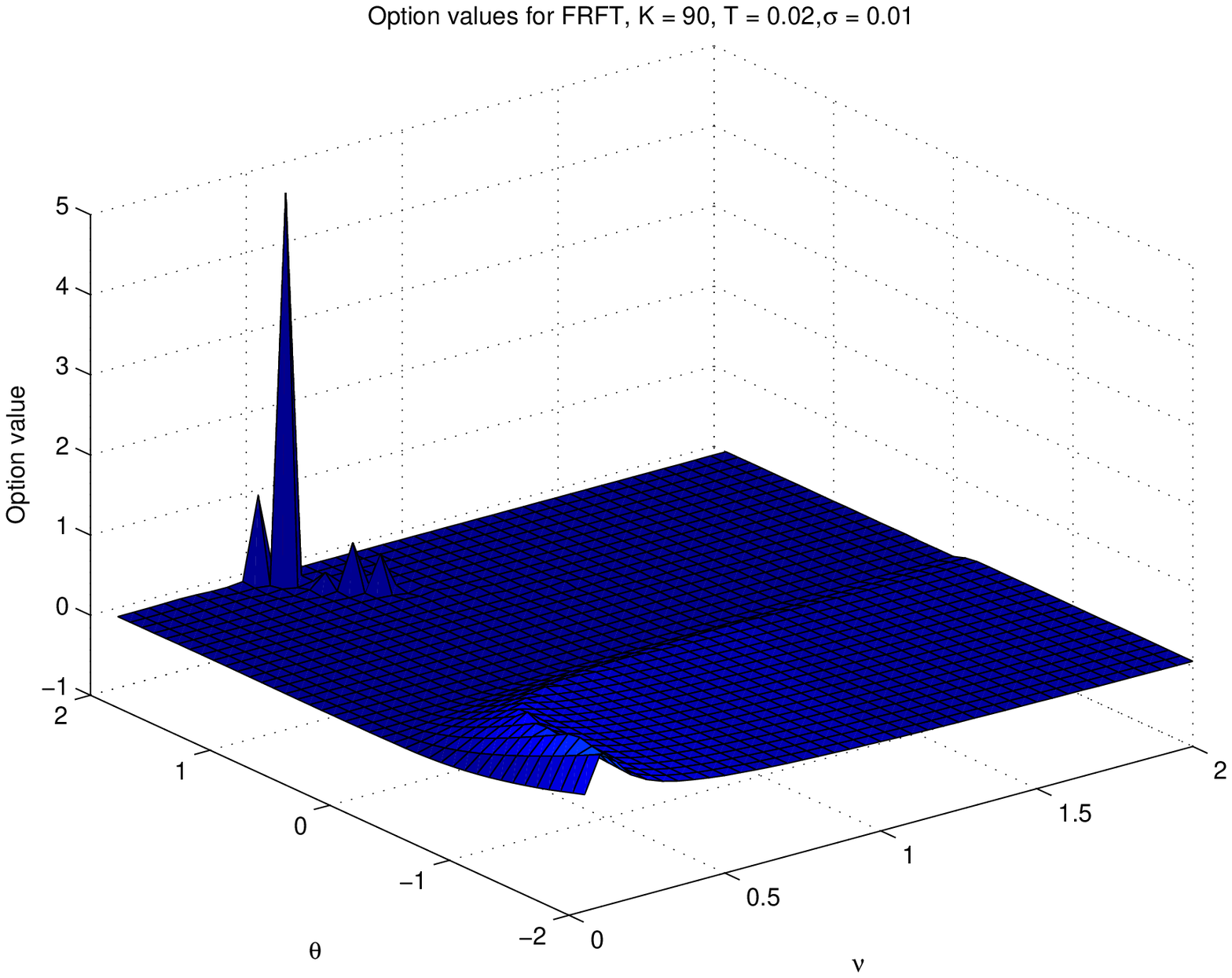}
\caption{European option values in VG model at $T=0.02 yrs, K = 90, \sigma = 0.01$
obtained with FRFT.}
\label{MikeFRFT1}
\end{minipage}
\hspace{0.1\textwidth}
\begin{minipage}[ht]{0.4\textwidth}
\includegraphics[totalheight=2in]{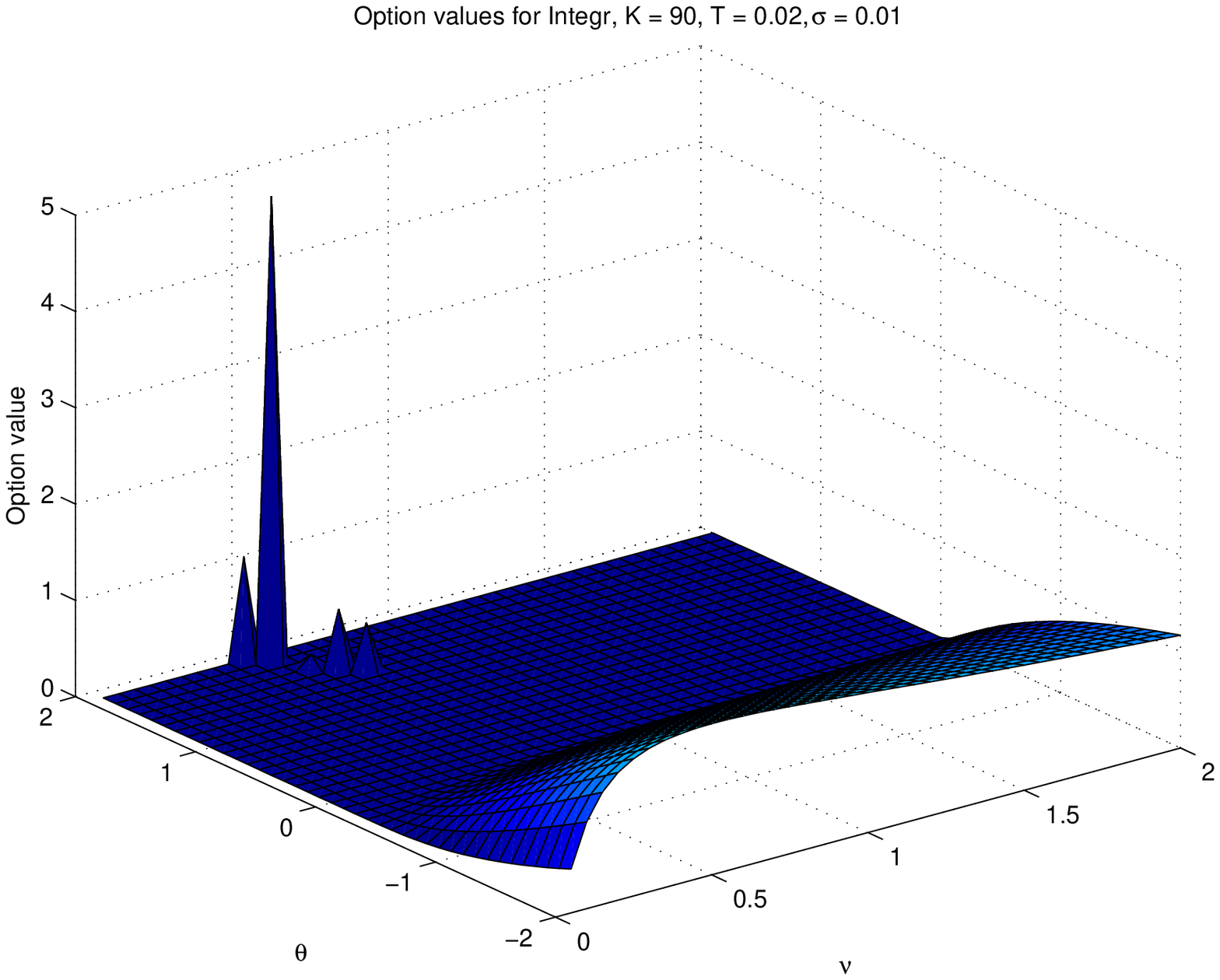}
\caption{European option values in VG model at $T=0.02 yrs, K = 90, \sigma = 0.01$
obtained with the adaptive integration.}
\label{MikeIntegr1}
\end{minipage}
\end{flushleft}
\end{figure}

Parameter $\alpha $ in Eq.~(\ref{callFFT}) must be positive. Usually
$\alpha
 = 3$ works well for various models. It is important that the denominator in
 Eq.~(\ref{omega}) has only imaginary roots while integration in
 Eq.~(\ref{callFFT}) is provided along real $v$. Thus, the integrand of
 Eq.~(\ref{callFFT}) is well-behaved.

But as it turned out, this is not the case for the VG model. To show
this let us consider the European call option values obtained by
Mike Konikov by computing FFT of the VG characteristic function
according to Eq.~(\ref{callFFT}).

\begin{figure}[ht]
\begin{flushleft}
\begin{minipage}[ht]{0.4\textwidth}
\includegraphics[totalheight=2in]{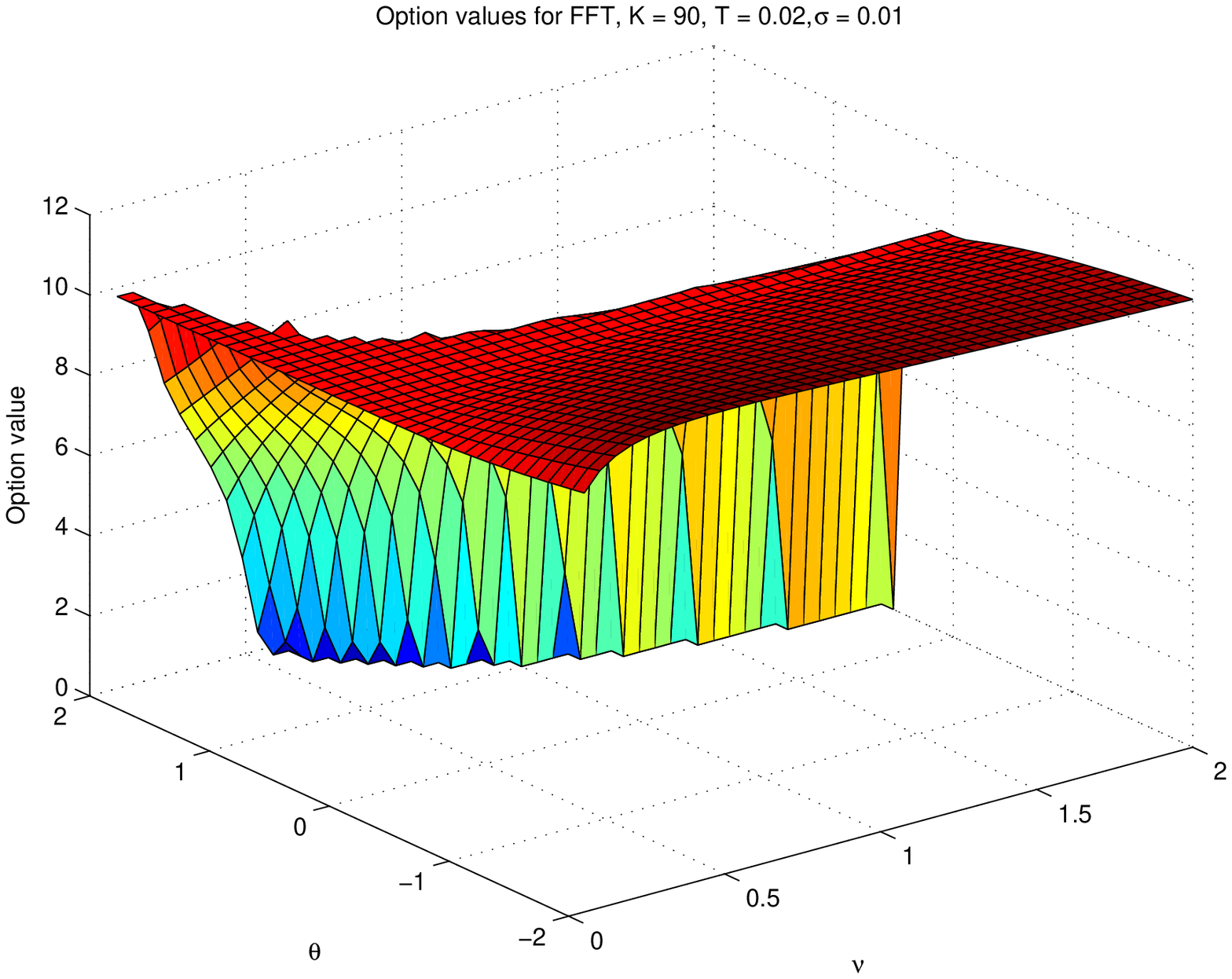}
\caption{European option values in VG model at $T=0.02 yrs, K = 90, \sigma = 0.01$
obtained with FFT.}
\label{MikeFFT1}
\end{minipage}
\hspace{0.1\textwidth}
\begin{minipage}[ht]{0.4\textwidth}
\includegraphics[totalheight=2in]{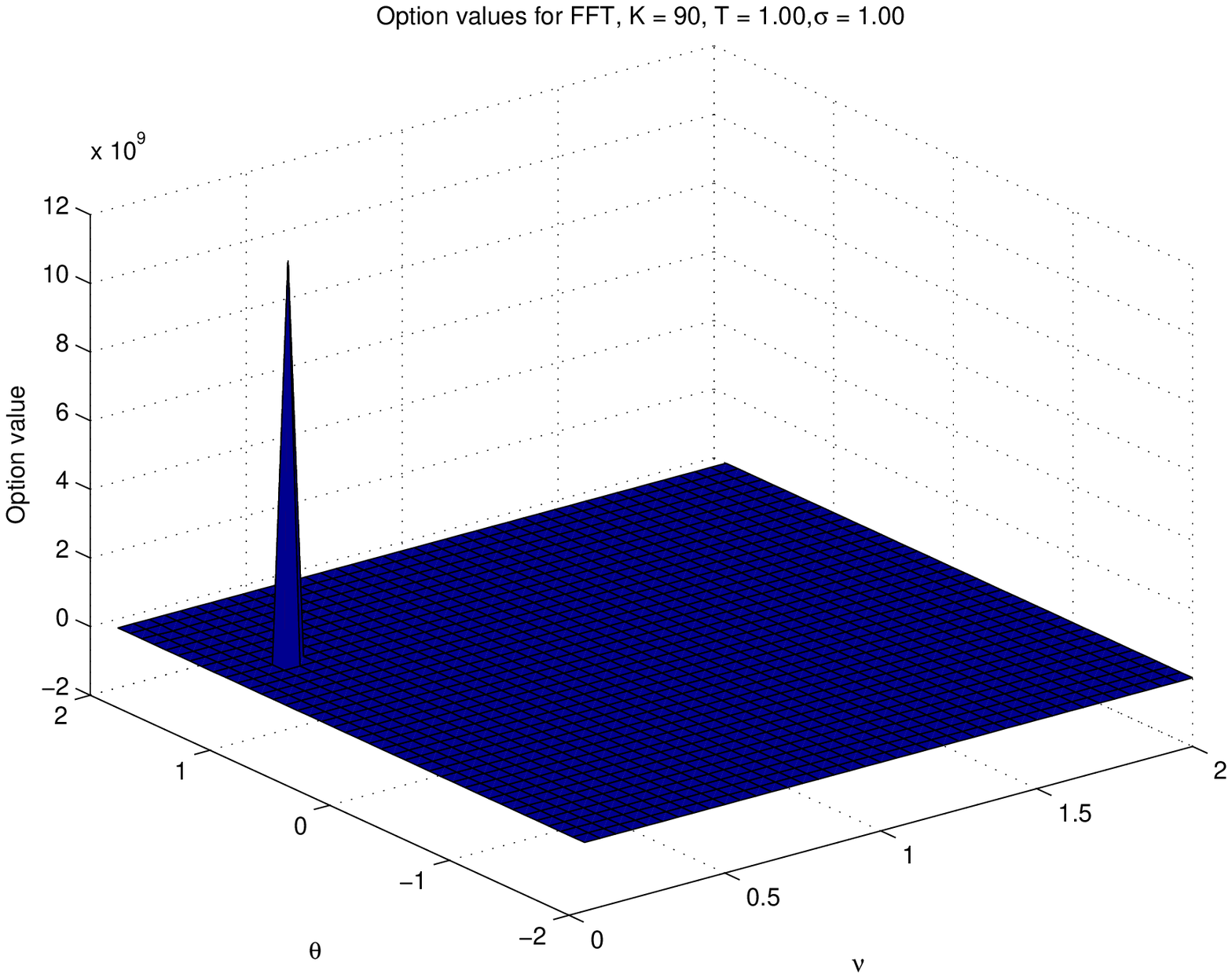}
\caption{European option values in VG model at $T=1.0 yrs, K = 90, \sigma = 1.0$
obtained with the FFT.}
\label{MikeFFT2}
\end{minipage}
\end{flushleft}
\end{figure}

\begin{figure}[ht]
\begin{flushleft}
\begin{minipage}[ht]{0.4\textwidth}
\includegraphics[totalheight=2in]{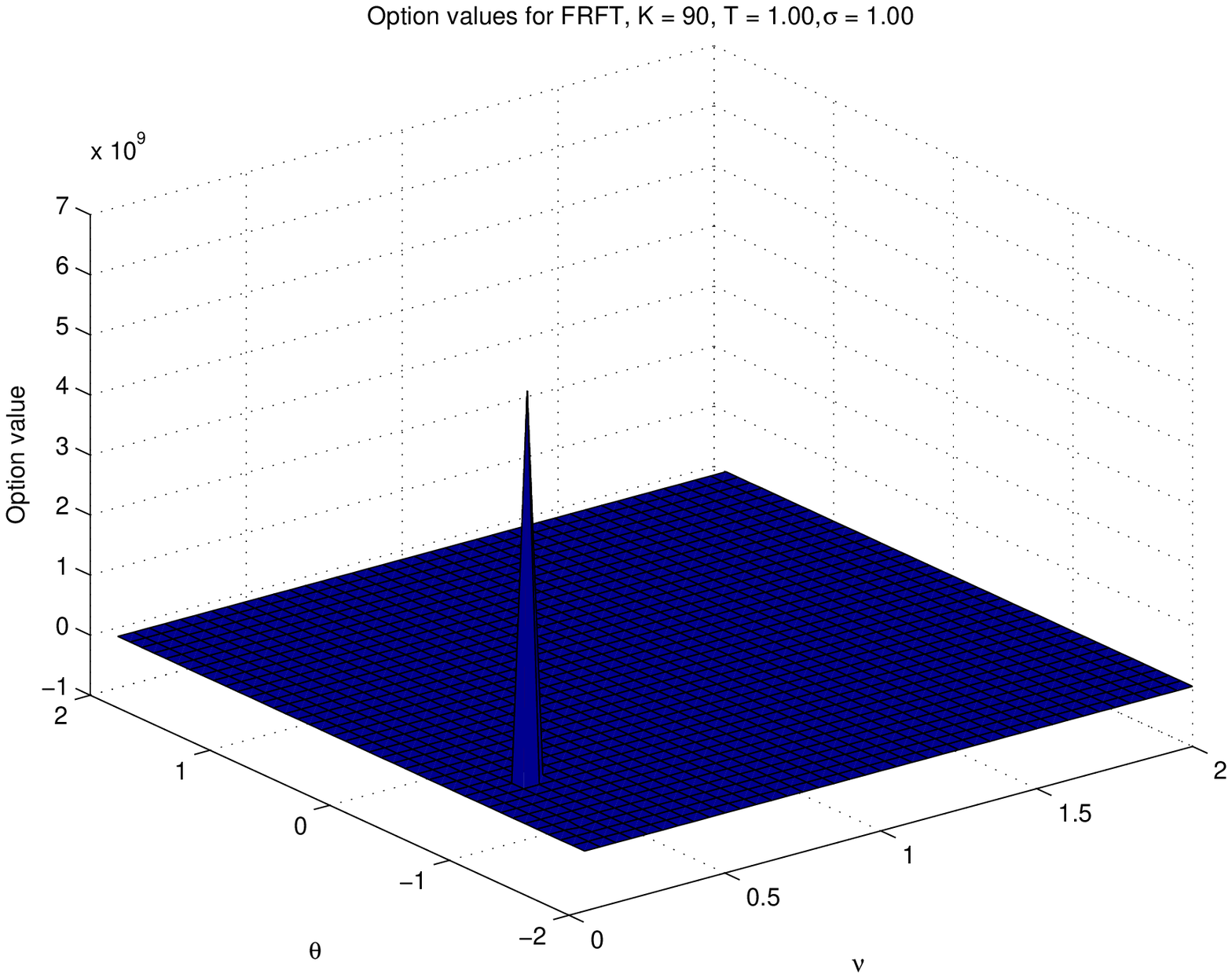}
\caption{European option values in VG model at $T=1.0 yrs, K = 90, \sigma = 1.0$
obtained with the FRFT.}
\label{MikeFRFT2}
\end{minipage}
\hspace{0.1\textwidth}
\begin{minipage}[ht]{0.4\textwidth}
\includegraphics[totalheight=2in]{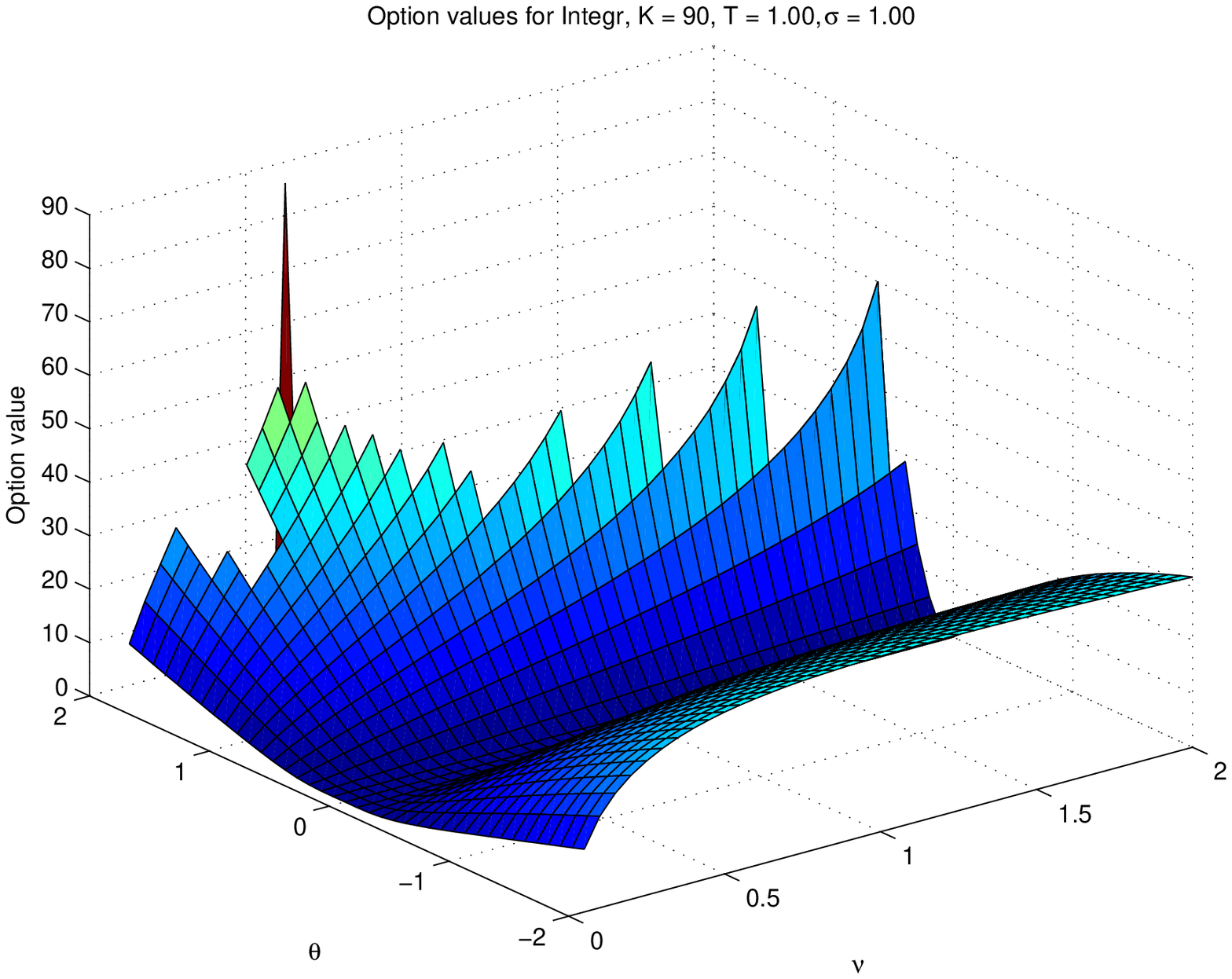}
\caption{European option values in VG model at $T=1.0 yrs, K = 90, \sigma = 1.0$
obtained with the adaptive integration.}
\label{MikeIntegr2}
\end{minipage}
\end{flushleft}
\end{figure}

In Fig.~\ref{MikeFRFT1} the results of that test obtained using the FRFT
algorithm are given for strike $K=90$, maturity $T = 0.02$ yrs and volatility
$\sigma = 0.01$. It is seen that at positive coefficients of skew $\Theta
\approx 2$ and coefficients of kurtosis $\nu \approx 0.5$ the option value has
a delta-function-wise pick that doesn't seem to be a real option value
behavior. In Fig.~\ref{MikeIntegr1} similar results are obtained using a
different method of evaluation of the integral in Eq.~(\ref{callFFT}) - an
adaptive integration. Eventually, in Fig.~\ref{MikeFFT1} same test was provided
using a standard FFT method. The results look quite different that allows a
guess that something is wrong with FRFT and the adaptive integration. One could
also note that this test plays with an option with a very short maturity.
Therefore, to let us make another test with a longer maturity. In
Fig.~\ref{MikeFFT2}-\ref{MikeIntegr2} the results of the test that uses
same integration procedures, but for the option with $K = 90, T=1, \sigma = 1$,
are presented. It is seen that for longer maturities FFT also blows up almost
at the same region of the model parameters. Moreover, it occurs not only at
positive value of the skew coefficient but at negative as well. Thus, the
problem lies not in the numerical method that was used to evaluate the integral
in the Eq.~(\ref{callFFT}), but in the integral itself.

Now having expression Eq.~(\ref{char1}) for the VG characteristic
function let us substitute it and  Eq.~(\ref{omega}) into the
Eq.~(\ref{callFFT}) that gives

\begin{equation} \label{callFFT1}
C(K,T) \propto \dfrac{e^{-\alpha\log(K) - rT}}{\pi}\int_0^{\infty}
\Re\left\{\dfrac{e^{-iv\log(K)}}{\left[\alpha^2+\alpha-v^2+i(2\alpha+1)v
\right]\left( 1 - i\theta\nu u + \frac{1}{2}\sigma^{2}\nu
u^{2}\right)^{\frac{t}{\nu}} }\right\}dv,
\end{equation}

\begin{figure}[ht]
\begin{flushleft}
\begin{minipage}[ht]{0.4\textwidth}
\includegraphics[totalheight=2in]{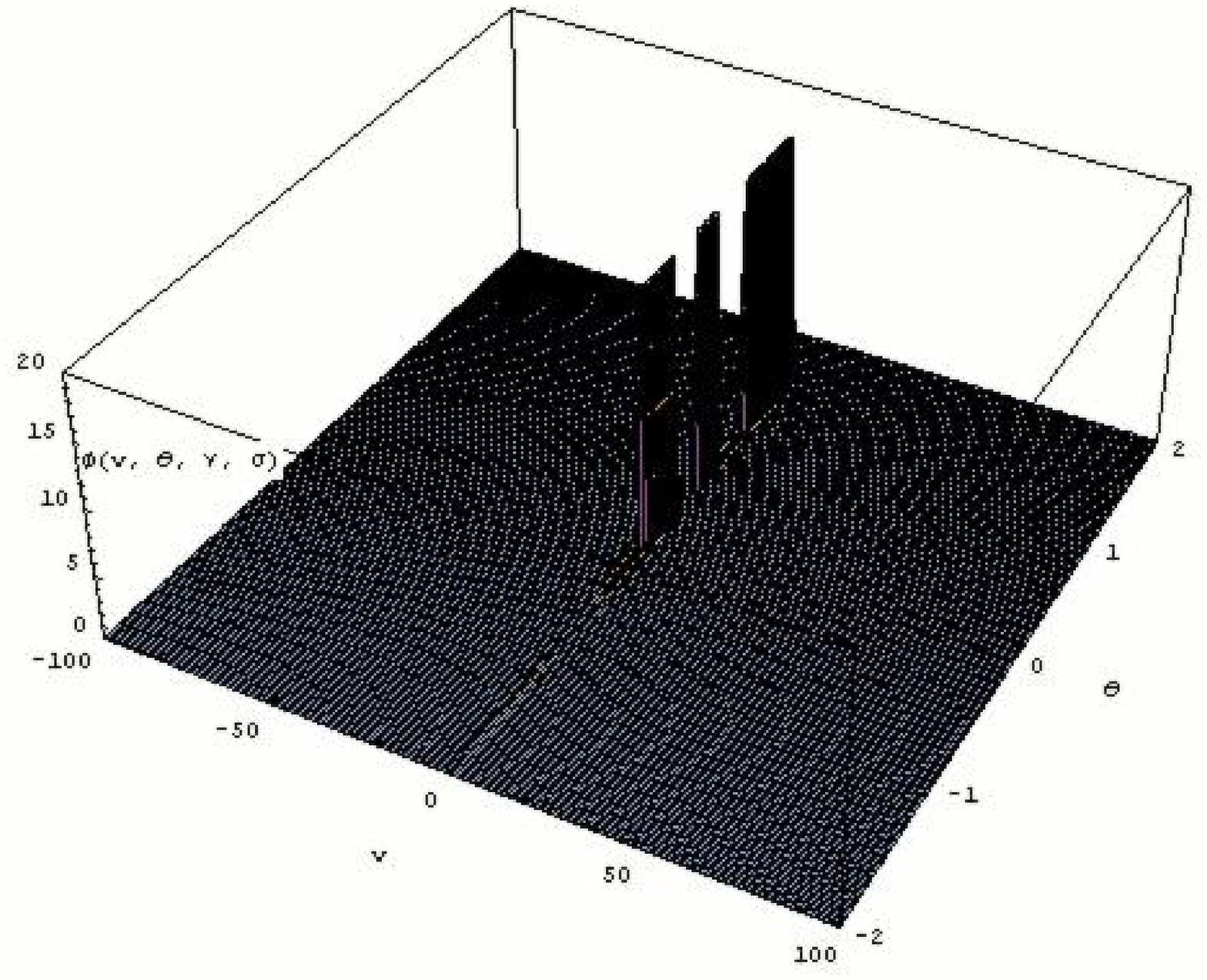}
\caption{Denominator of the  Eq.~(\ref{callFFT1}) at $T = 0.8, \nu = 0.1,
\alpha =3, \sigma = 1$ as a function of $v$ and $\Theta $.}
\label{Math1}
\end{minipage}
\hspace{0.1\textwidth}
\begin{minipage}[ht]{0.4\textwidth}
\includegraphics[totalheight=2in]{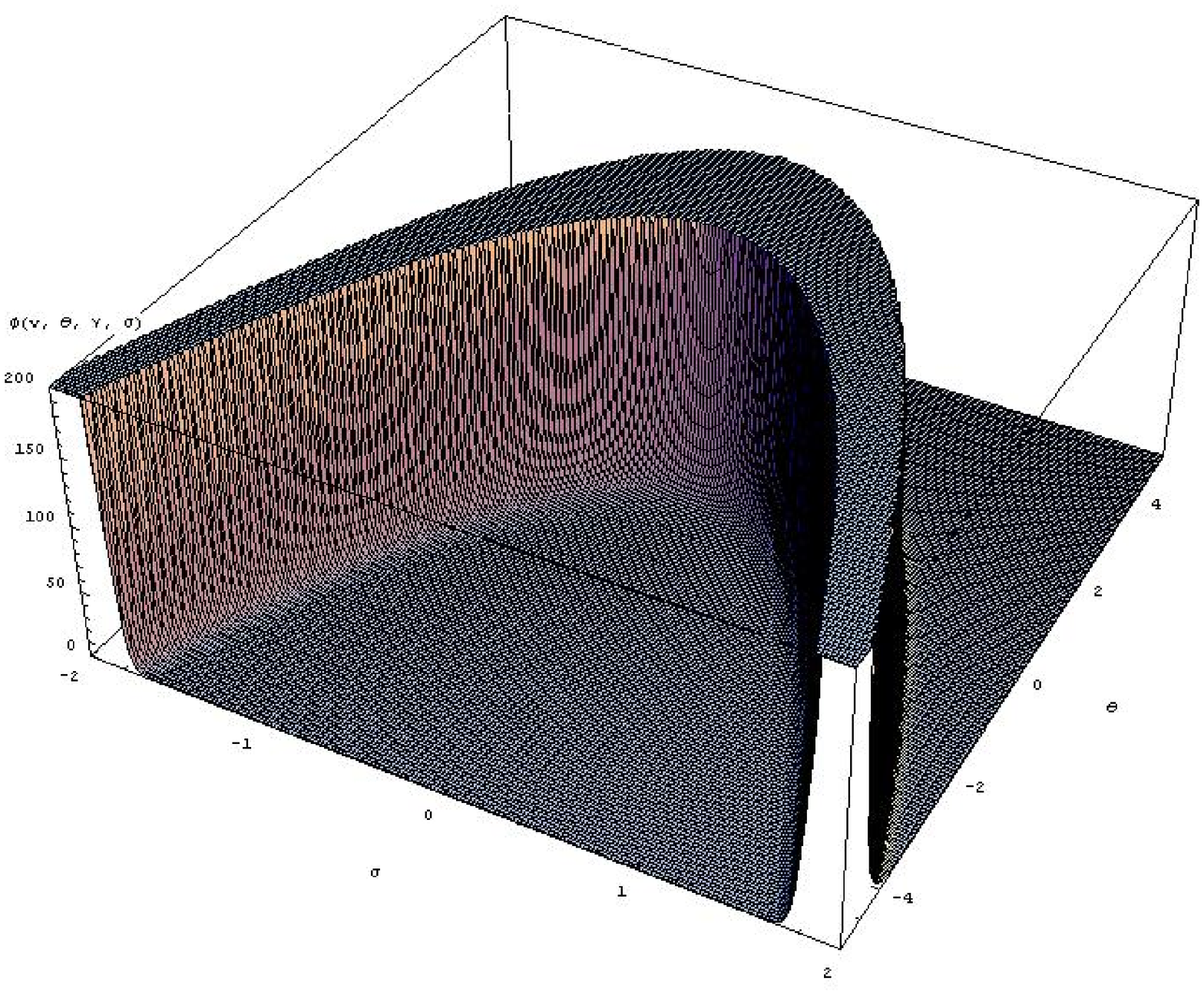}
\caption{Denominator of the  Eq.~(\ref{callFFT1}) at $T = 0.8, \nu = 0.1,
\alpha =3, v=0$  as a function of $\sigma$ and $\Theta $.} \label{Math2}
\end{minipage}
\end{flushleft}
\end{figure}

\ni where $u \equiv v-(\alpha+1)i$. At small $T$ close to zero the second term
in the denominator of the Eq.~(\ref{callFFT1}) is close to 1. Therefore at
small $T$ the denominator has no real roots. To understand what happens at
larger maturities, let us put $T = 0.8, \nu = 0.1, \alpha =3, \sigma = 1$ and
see how the denominator behaves as a function of $v$ and $\Theta $. The results
of this test obtained with the help of Mathematica package are given in
Fig~\ref{Math1}.

It is seen that at $v=0$ at positive $\Theta $ the characteristic function has
a singularity. To investigate it in more detail, we assume $v=0$ and plot the
denominator as a function of $\sigma $ and $\Theta$ (see Fig.~\ref{Math2}). As
follows from this Figure in the interval $0 < \sigma < 2$ there exists a value
of $\Theta $ that makes the integrand in the Eq.~(\ref{callFFT1}) singular.
This means that singularity of the integrand can not be eliminated, and thus
the Carr-Madan FFT method can not be used together with the VG model for
pricing European vanilla options. Using FRFT or adaptive integration that both
are slight modifications of the FFT, also doesn't help.

Note that for the VG model the authors of \cite{CarrMadan:99a} derived
condition which keeps the characteristic function to be finite, that reads

\begin{equation}\label{cond}
  \alpha  < \sqrt{\dfrac{2}{\nu \sigma ^2} + \dfrac{\Theta ^2}{\sigma ^4}} -
\dfrac{\Theta }{\sigma ^2} - 1.
\end{equation}

Also as can be seen, for $\Theta, \nu $ and $\sigma $ corresponding to the
above mentioned tests $\alpha $ becomes negative that doesn't allow using this
method to price the options in terms of strike.

In order to solve these problems one needs to find another way how to
regularize the integrand, i.e. eliminate doing it in the way as Carr and Madan
did it using a regularization factor $e^{-\alpha k}$.

\section{Lewis's regularization}
Another approach of how to apply FFT to the pricing of European options was
proposed by Alan Lewis \cite{Lewis:2001}. Lewis notes that a general integral
representation of the European call option value with a vanilla payoff is

\begin{equation}\label{genInt}
  C_T(x_0, K) = e^{-rT} \int_{-\infty}^{\infty} \left( e^x - K\right)^+ q(x, x_0,
  T)dx,
\end{equation}

\ni where $x = \log S_T$ is a stock price that under a pricing measure evolves
as $S_T = S_0\exp[(r-q)T + X_T$, $r-q$ is the cost of carry, $T$ is the
expiration time for some option, $X_T$ is some Levy process satisfying
$\E[exp(i u X_T)] =1$, and $q$ is the density of the log-return
distribution $x$.

The central point of the Lewis's work is to represent the
Eq.~(\ref{genInt}) as a convolution integral and then apply a Parseval
identity

\begin{equation}\label{parseval}
\int_{-\infty}^{\infty} f(x) g(x_0-x)dx = \dfrac{1}{2\pi}
\int_{-\infty}^{\infty} e^{-i u x_0}\hat{f}(u)\hat{g}(u)du,
\end{equation}
\ni where the hat over function denotes its Fourier transform.

The idea behind this formula is that the Fourier transform of a transition
probability density for a Levy process to reach $X_t = x$ after the elapse of
time $t$ is a well-known characteristic function, which plays an important role
in mathematical finance. For Levy processes it is $\phi _t(u) =
\E[\exp(iuX_t)], u \in \Re$, and typically has an analytic extension (a
generalized Fourier transform) $u \rightarrow z \in {\mathbb C}$, regular in
some strip ${\cal S}_X$ parallel to the real z-axis.

Now suppose that the generalized Fourier transform of the payoff
function $\hat{w}(z) = \int_{-\infty}^{\infty} e^{izx}(e^x - K)^+dx$
and the characteristic function $\phi _t(z)$ both exist (we will
discuss this below). Then from a chain of equalities the call option
value can be expressed as follows

\begin{eqnarray}  \label{chain}
C_T(x_0, K) &=& e^{-rT} \E\left[ \left( e^x - K\right)^+\right]
    = \dfrac{e^{-rT}}{2\pi}\E \left[\int_{i\mu -\infty}^{i\mu +\infty}
    e^{-izx_T} \hat{w}(z)dz \right]
\\
&=& \dfrac{e^{-rT}}{2\pi}\E \left[\int_{i\mu -\infty}^{i\mu +\infty} e^{-iz[x_0
+ (r-q + \omega)T]} e^{-izX_T} \hat{w}(z)dz \right]
\nn \\
 &=& \dfrac{e^{-rT}}{2\pi}\int_{i\mu -\infty}^{i\mu +\infty}
 e^{-iz[x_0 + (r-q+ \omega)T]} \E[e^{-izX_T}] \hat{w}(z)dz =
\dfrac{e^{-rT}}{2\pi}\int_{i\mu -\infty}^{i\mu +\infty}
 e^{-izY} \phi_{X_T} (-z) \hat{w}(z)dz.
 \nn
\end{eqnarray}

Here $Y = x_0 + (r-q+ \omega)T$, $\mu \equiv$ Im $z$. This is a formal
derivation which becomes a valid proof if all the integrals in
Eq.~(\ref{chain}) exist.

The Fourier transform of the vanilla payoff can be easily found by a direct
integration

\begin{equation}\label{FTpayoff}
\hat{w}(z) = \int_{-\infty}^{\infty} e^{izx}(e^x - K)^+dx = -
\dfrac{K^{iz+1}}{z^2 - iz}, \qquad \mathrm {Im} z > 1.
\end{equation}

Note that if z were real, this regular Fourier transform would not exist. As
shown in \cite{Lewis:2000}, payoff transforms $\hat{w}(z)$ for typical claims
exist and are regular in their own strips ${\cal S}_w$ in the complex z-plane,
just like characteristic functions.

Above we denoted the strip where the characteristic function $\phi (z)$ is
well-behaved as ${\cal S}_X$. Therefore, $\phi (-z)$ is defined at the
conjugate strip ${\cal S}^*_X$. Thus, the Eq.~(\ref{chain}) is defined at the
strip ${\cal S}_V = {\cal S}^*_X \bigcap {\cal S}_w$, where it has the form

\begin{equation}\label{callFFTfin}
C(S,K,T) = - \dfrac{Ke^{-rT}}{2\pi}\int_{i\mu -\infty}^{i\mu +\infty}
 e^{-izk} \phi_{X_T} (-z) \dfrac{dz}{z^2-iz}, \quad \mu \in {\cal S}_V,
\end{equation}

\ni and $k = \log(S/K) + (r-q+ \omega)T$.

The characteristic function of the VG process has been given by the
Eq.~(\ref{charS}) and is defined in the strip $\beta - \gamma  <$ Im
$z < \beta + \gamma $, where

\begin{equation} \label{beta}
  \beta  = \dfrac{\Theta }{\sigma ^2}, \quad
  \gamma  = \sqrt{\dfrac{2}{\nu \sigma ^2} + \dfrac{\Theta ^2}{\sigma ^4} +
  2(\mathrm {Re} z)^2}.
\end{equation}

This condition can be relaxed by assuming in the Eq.~(\ref{beta}) $\mathrm {Re}
z = 0$ \footnote{In other words, if it is valid at $\mathrm {Re} z = 0$, it
will be valid for any $\mathrm {Re} z$}. Accordingly, $\phi (-z)$ is defined in
the strip $\gamma - \beta >$ Im $z > - \beta - \gamma $.

Now let us choose Im $z$ in the form

\begin{equation}\label{ImzForm}
\mu \equiv \mathrm{Im} \ z = \sqrt{1 + \dfrac{2\Theta }{\sigma ^2} +
\dfrac{\Theta ^2}{\sigma ^4}} - \dfrac{\Theta }{\sigma ^2}.
\end{equation}

Taking into account the Eq.~(\ref{constrain}) which makes a constrain on the
available values of the VG parameters, it is easy to see that $\mu $ defined in
such a way obeys the inequality $\mu < \gamma  - \beta $. On the other hand, as
also can be easily seen, $\mu \ge 1$ at any value of $\Theta $ and positive
volatilities $\sigma$, and the equality is reached when $\Theta =0$. It means,
that Im $z = \mu$ lies in the strip ${\cal S}^*_X$ as well as in  the strip
${\cal S}_w$, i. e. $\mu  \in {\cal S}_V$.

Now one more trick with contour integration. The integrand in
Eq.~(\ref{callFFTfin}) is regular throughout ${\cal S}^*_X$ except for simple
poles at $z = 0$ and $z = i$. The pole at $z = 0$ has a residue
$-Ke^{-rT}i/(2\pi)$, and the pole at $z = i$ has a residue $Se^{-qT}i/(2\pi)$
\footnote{This is because $\phi _T(-i) = e^{-\omega T}$}. The analysis of the
previous paragraph shows that the strip ${\cal S}^*_X$ is defined by the
condition $\gamma - \beta
> \mathrm{Im} z > - \beta - \gamma$, where $\gamma - \beta > 1$, and $- \beta -
\gamma < 0$. Therefore we can move the integration contour to $\mu_1 \in
(0,1)$. Then by the residue theorem, the call option value must also equal the
integral along Im $z = \mu_1$ minus $2\pi i$ times the residue at $z=i$. That
gives us a first alternative formula
\begin{equation} \label{altFFT}
C(S,K,T) =
    Se^{-qT} - \dfrac{Ke^{-rT}}{2\pi}\int_{i\mu_1
    -\infty}^{i\mu_1 +\infty}  e^{-izk} \phi_{X_T} (-z) \dfrac{dz}{z^2-iz}
\end{equation}

For example, with $\mu _1 = 1/2$ which is symmetrically located between the two
poles, this last formula becomes

\begin{equation} \label{altFFT2}
C(S,K,T) =
     Se^{-qT} - \dfrac{1}{\pi}\sqrt{SK}e^{-(r+q)T/2}
     \int_{0}^{\infty} \mathrm{Re}\left[e^{-iu \kappa } \Phi \left(-u -\dfrac{i}{2}
    \right) \right]\dfrac{du}{u^2+ \frac{1}{4}} \\
\end{equation}

\ni where $\kappa = \ln (S/K) + (r-q)T, \quad \Phi(u) = e^{i u \omega T} \phi_{X_T}(u)$ and
it is taken into account that the integrand is an even function of
its real part. The last integral can be rewritten in the form

\begin{equation} \label{intF}
\int_{0}^{\infty} e^{-iu \ln \kappa } \phi_1(u) du, \qquad
\phi _1(u) = \dfrac{4}{4u^2+ 1}\Phi\left(-u - \dfrac{i}{2}\right) .
\end{equation}

This can be immediately recognized as a standard inverse Fourier
transform, and by derivation the integrand is regular everywhere.
Indeed, $\phi _{X_T}(-u-i/2)|_{u=0} = (1-\frac{\sigma ^2 \nu }{8} -
\frac{\nu \theta }{2})^{-t/\nu}$, therefore the denominator vanishes
if $\frac{2}{\nu}  = \theta + \frac{\sigma ^2}{4}$. Now using the
Eq.~(\ref{constrain}) one finds that $\theta + \frac{\sigma ^2}{4} >
2h + \sigma ^2$ or $\frac{\sigma ^2}{4} < -\frac{\theta }{3}$. Thus,
$\theta $ must be negative to turn the denominator to zero. The last
equality could be also rewritten as $\theta +\frac{\sigma ^2}{4} <
\frac{2\theta }{3}$. Thus, the denominator vanishes if $\frac{1}{\nu } < \frac{2\theta }{3}$,
i.e. $\nu $ must be negative, but it is not! Therefore, the characteristic function in Eq.~(\ref{intF})
doesn't have singularity at $u=0$. Thus, a standard FFT or FRFT method can be applied to get the value
of the integral.

\begin{figure}[bht]
\begin{flushleft}
\begin{minipage}[ht]{0.4\textwidth}
\includegraphics[totalheight=2in]{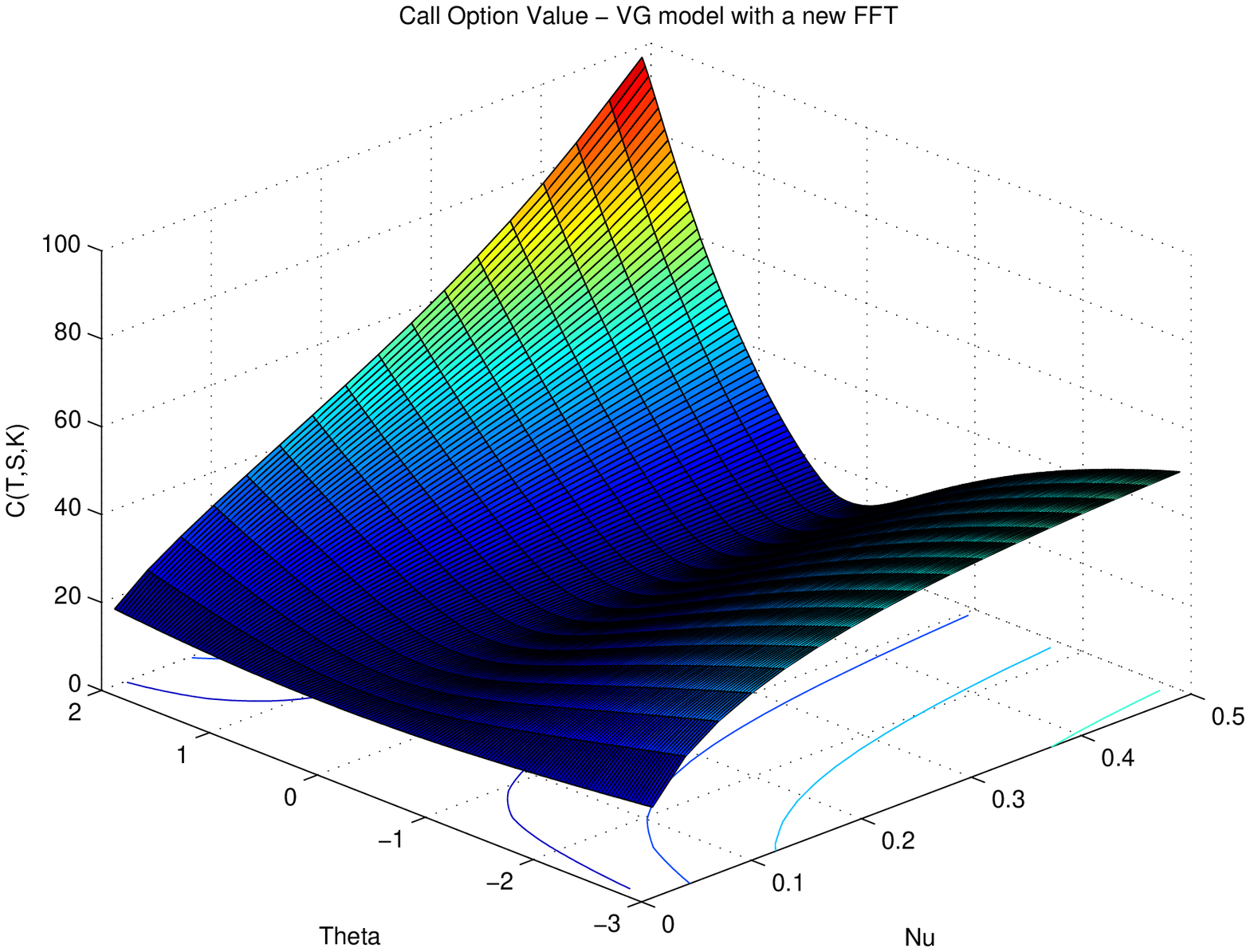}
\caption{European option values in VG model at $T=1.0 yr, K = 90, \sigma = 0.1$
obtained with the new FFT method.} \label{MyFFT1}
\end{minipage}
\hspace{0.1\textwidth}
\begin{minipage}[ht]{0.4\textwidth}
\includegraphics[totalheight=2in]{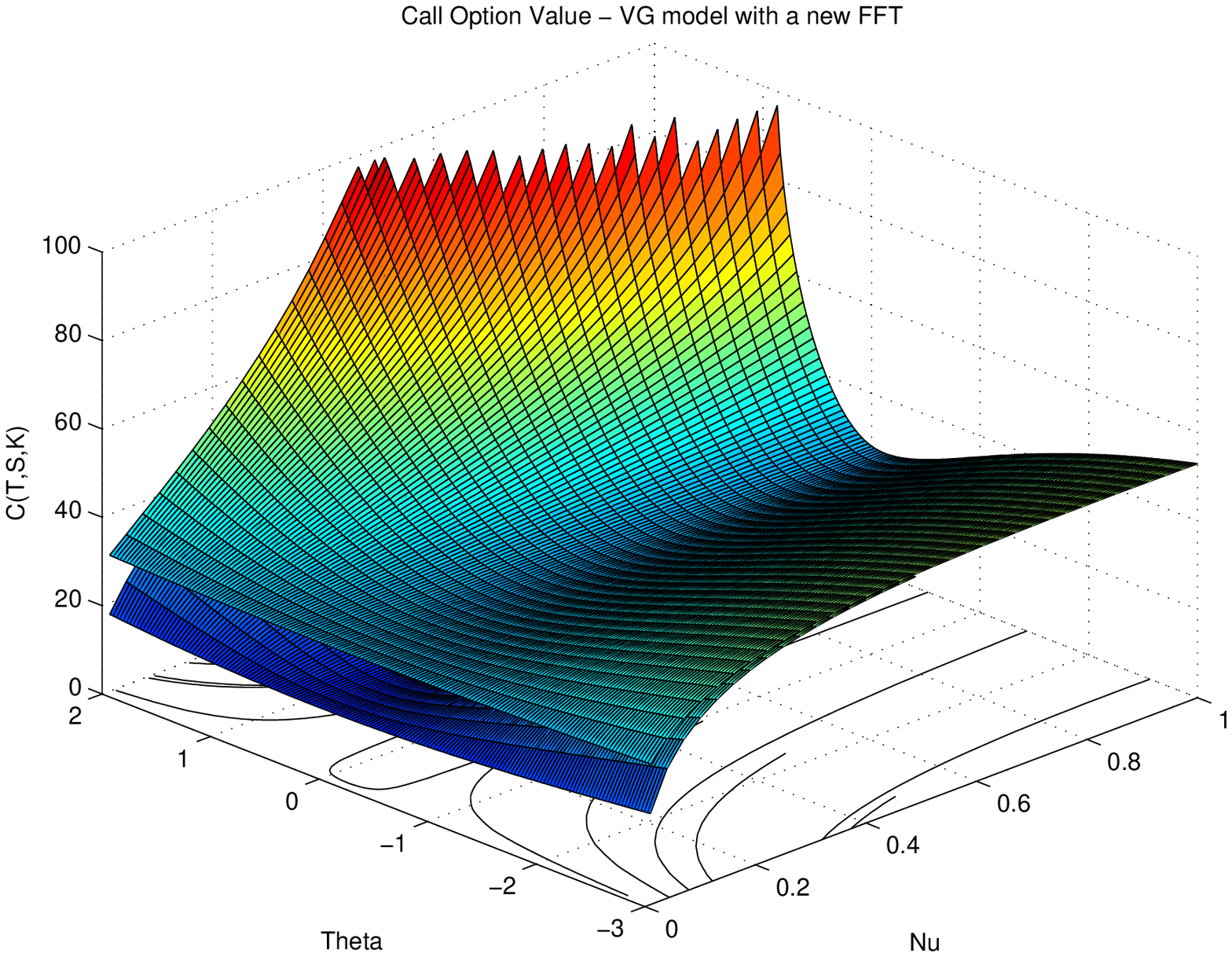}
\caption{European option values in VG model at $T=1.0 yrs, K = 90, \sigma =
0.5$ obtained with the new FFT method.} \label{MyFFT2}
\end{minipage}

\end{flushleft}
\begin{center}
\includegraphics[totalheight=2in]{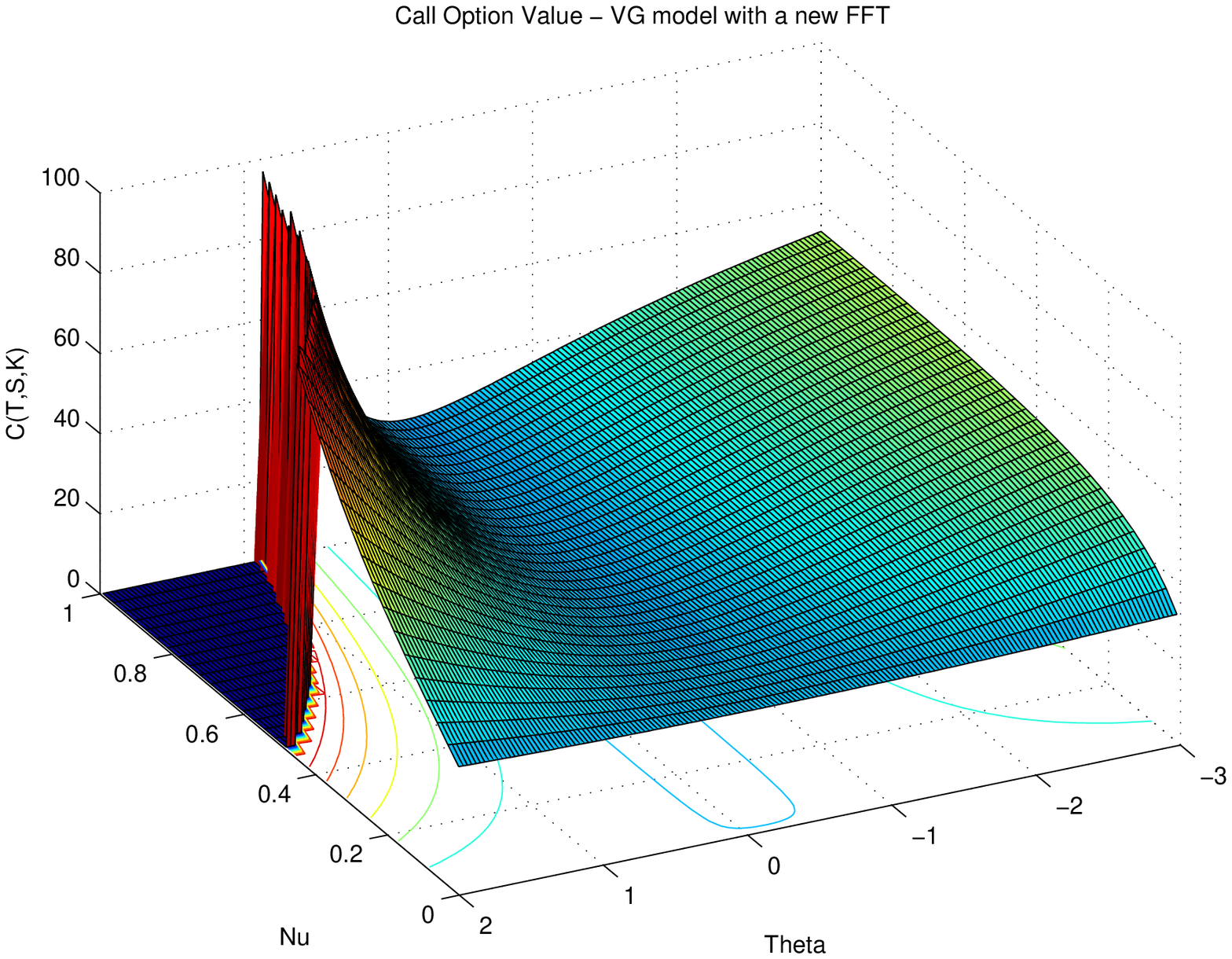}
\caption{European option values in VG model at $T=1.0 yr, K = 90, \sigma = 0.5$
obtained with the new FFT method (rotated graph).} \label{MyFFT2_2}
\end{center}
\end{figure}

In Fig.~\ref{MyFFT1} -\ref{MyFFT2} the results of the European vanilla option
pricing with the VG model conducted by using this new FFT method are displayed.
Two test has been provided with parameters $T=1$ yr, $K=90, \sigma = 0.1$
(Fig.~\ref{MyFFT1}) and $T=1$ yr, $K=90, \sigma = 0.5$ (Fig.~\ref{MyFFT2}). It
is seen that the option value surface is regular in both cases. Zero values
indicates that region, where the VG constrain Eq.~(\ref{constrain}) is not
respected. The higher values of $\sigma $ and $\Theta$ are the lower values of
$\nu $ are required to obey this constraint. Therefore, at higher values of
$\nu $ the model is not defined that produces irregularity in the graph. This
effect is better observable in Fig.~\ref{MyFFT2_2} that is obtained by rotation
of the Fig.~\ref{MyFFT2}. The above means that the new FFT method can be used
with no essential problem. A generalization of this method for FRFT is also
straightforward.

\begin{figure}[bht]
\begin{center}
\includegraphics[totalheight=1.5in]{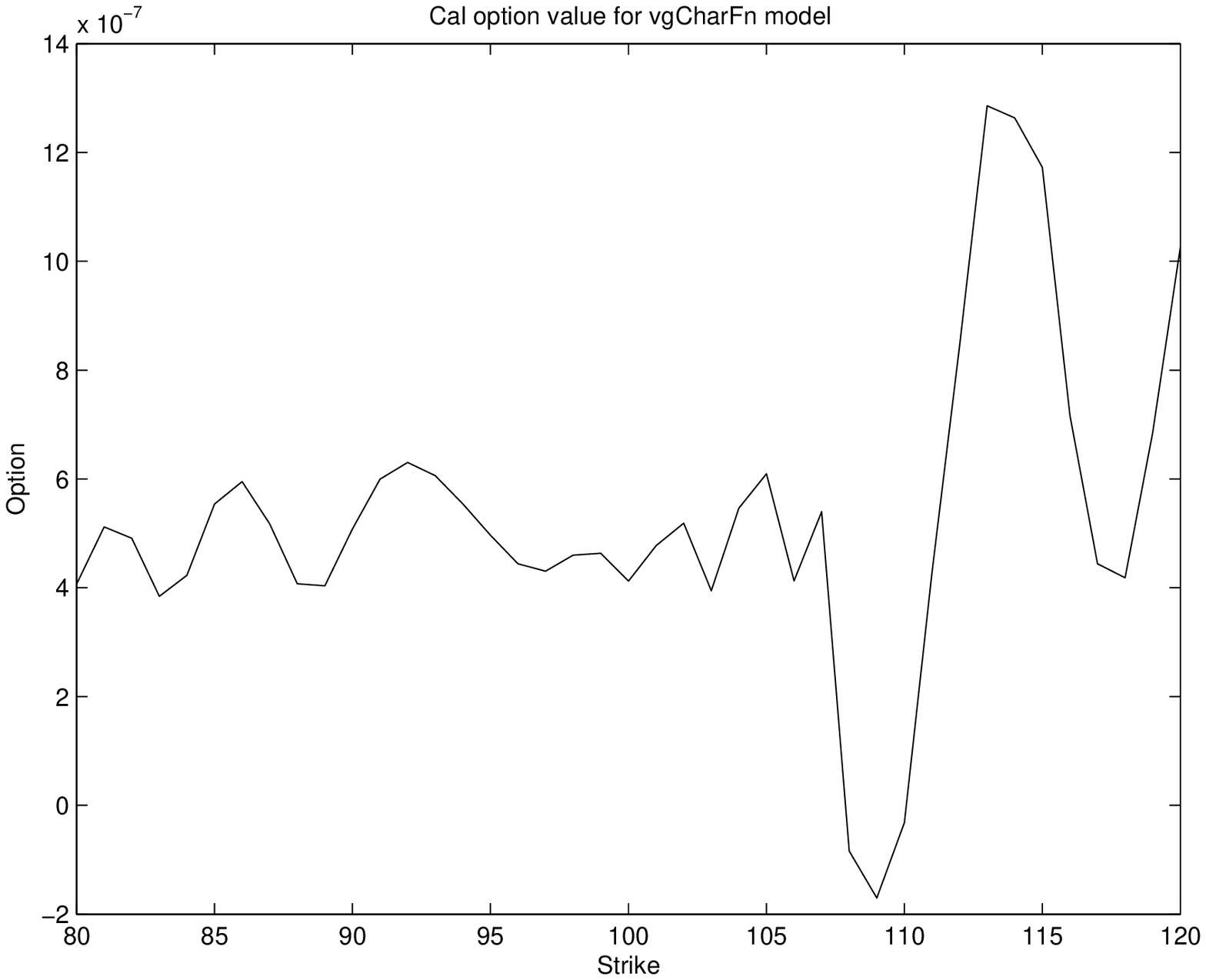}
\caption{The difference between the European call option values for the VG
model obtained with Carr-Madan FFT method and the new FFT method. Parameters of
the test are: $S=100, T=0.5 yr, \sigma = 0.2, \nu =0.1, \Theta =-0.33, r=q=0.$
at various strikes).} \label{diff}
\end{center}
\end{figure}

In the region of the VG parameters values where an application of the
Carr-Madan FFT procedure doesn't cause the problem the results of that
method are almost identical to what the described above method gives.
An example of such a comparison is given in Fig.~\ref{diff} (my NewFFT
Matlab code vs Mike's FFT code). It is seen that the difference is of
the order of $10^{-7}$.

\section{Black-Scholes-wise method}

One more method of regularization of the Fourier kernel for the VG
model has been proposed by Sepp  \cite{Sepp2003} and is also discussed in \cite{Iddo2004VG}, \cite{ContTankov2004}. The idea is
as follows.

Given characteristic function $\phi_{X_t} (z)$ of the model $M$ the
price of a European option can be expressed as
\begin{eqnarray}\label{HEuropPrice}
\Pi_1^M &=& \dfrac{1}{2} + \dfrac{\xi}{2\pi} \int_{-\infty}^{\infty}
\dfrac{e^{-iu \ln K} e^{iu [\ln S +(r-q+\omega)T]} \phi_{X_T}(u-i)}{i u \phi_{X_T}(-i)}du, \\
\Pi_2^M &=& \dfrac{1}{2} + \dfrac{\xi}{2\pi} \int_{-\infty}^{\infty}
\dfrac{e^{-iu \ln K}e^{iu [\ln S +(r-q+\omega)T]}\phi_{X_T}(u)}{i u}du, \nn \\
V^M &=& \xi \left[e^{-q T}S_0 \Pi_1^M - e^{-r T} K \Pi_2^M\right]
\nn,
\end{eqnarray}

\ni where $\xi = 1(-1)$ for a call(put). Eq.~(\ref{HEuropPrice}) is
a generalization of the Black-Scholes option pricing formula. Note
that $\phi_{X_t} (0) = 1$ by definition, and $\phi_{X_t} (-i)$ is a
function of time to expiry $T$ and parameters of the model only.

{\bf Proof}: Assume that $\phi _T(-z)$ has a strip of regularity $0
\le \mu \le 1$. First we rewrite Eq.~(\ref{callFFTfin}) as

\begin{eqnarray}  \label{BSFFT}
C(S,K,T) &=& - \dfrac{K e^{-rT}}{2\pi}\int_{i\mu -\infty}^{i\mu
    +\infty} e^{-izk}\phi_{X_T} (-z) \dfrac{dz}{z^2-i z} \\
&=&   - \dfrac{Ke^{-rT}}{2\pi}\left[\int_{i\mu -\infty}^{i\mu
    +\infty} e^{-izk} \phi_{X_T} (-z) \dfrac{i dz}{z} - \int_{i\mu -\infty}^{i\mu
    +\infty} e^{-izk} \phi_{X_T}(-z) \dfrac{i dz}{z-i} \right]\nn \\
 &=& - \dfrac{Ke^{-rT}}{2\pi}({\cal R}(I_1) - {\cal R}(I_2)) \nn
\end{eqnarray}

\begin{figure}[ht]
\begin{center}
\includegraphics[totalheight=2 in]{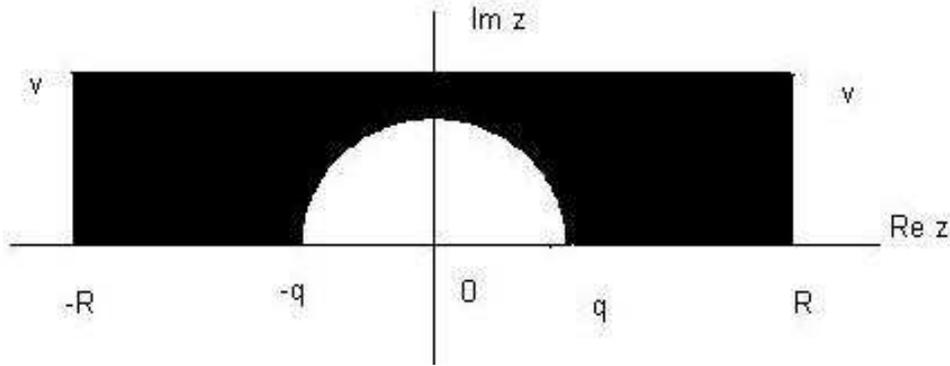}
\caption{Integration contour for ${\cal R}(I_1)$}. \label{Contour}
\end{center}
\end{figure}

In order to evaluate $I_1$ we employ a contour integral over the
contour given by 6 parametric curves (see Fig.~(\ref{Contour}):
$\Gamma_1: z=u, u \in (q,R), q,R > 0; \Gamma_2 : z = R + ib, b \in
(0, v); \Gamma_3: z = u + iv, u \in (R,-R); \Gamma_4 : z = -R + ib,
b \in (v, 0); \Gamma_5 : z = u, u \in (-R,-q); \Gamma_6 : z =
qe^{i\theta }, \theta \in (\pi, 0)$. As the integrand is analytic on
this contour we can apply the Cauchy theorem. Also note that the
integral along curve $\Gamma_6$ is a half of the integral along the
whole circle around zero which in turn is equal to $2\pi i^2
Res(e^{-izk}\phi_t(-z)/z)$. As the integrals along vertical lines
vanish at $R \rightarrow \infty$ and at $q \rightarrow 0$ the
integral along the real axis tends to an integral from $-\infty$ to
$\infty$, eventually changing variable $u \rightarrow - u$  we
obtain

\begin{equation} \label{FirstInt}
  {\cal R}(I_1) = \pi + \int_{-\infty}^{\infty}e^{- iu \ln K }e^{iu [\ln S +(r-q+\omega)T]}\dfrac{\phi_{X_T}(u)}{iu}du.
\end{equation}

To compute the ${\cal R}(I_2)$ we use a similar contour build around
the point $z=i$, i.e. $\Gamma _1 : z = u + i, u \in (q,R), q,R
> 0; \Gamma _2 : z = R+ib, b \in (1, 1+v); \Gamma _3 : z = u+i(1+v), u \in (R,-R);
\Gamma _4 : z = -R+ib, b \in (v, 1); \Gamma _5 : z = u + i, u \in
(-R,-q); \Gamma _6 : z = i + qe^{i\theta}, \theta \in (0, \pi)$.
Again taking limits $R \rightarrow \infty$ and $q \rightarrow 0$,
changing variable $ u \rightarrow u-i$, we obtain

\begin{equation} \label{SecondInt}
  {\cal R}(I_2) = \dfrac{S}{K}e^{(r-q)T}\left( \pi + \int_{-\infty}^{\infty}e^{-iu \ln K}e^{iu [\ln S +(r-q+\omega)T]}
  \dfrac{\phi_{X_T}(u-i)}{iu \phi_{X_T}(-i)}du \right).
\end{equation}

Substituting these integrals into the Eq.~(\ref{BSFFT}) we obtain
the Eq.~(\ref{HEuropPrice}) $\blacksquare$.

The difficulty in using FFT to evaluate the
Eqs.~(\ref{HEuropPrice}), as noted by Carr and Madan is the
divergence of the integrands at $u=0$. Specifically, let us develop
the characteristic function $\phi_{X_t} (z)$ with $z = u +iv$ as
Taylor series in $u$

\begin{equation}\label{Taylor}
  \phi_{X_t} (z) = \E[e^{-v X_t}] + iu\E[x e^{-v X_t}] - \frac{1}{2}
  u^2 \E[x^2e^{-v X_t}] + ...
\end{equation}

In Eq.~(\ref{altFFT}) we have to chose $z=u-i$ in the first
expression, and $z=u$ in the second one. As it is easy to check in
both cases that the leading term in the expansion under both
integrals is $1/(iu)$ which is just a source of the divergence.The
source of this divergence is a discontinuity of the payoff function
at $K=S_T$. Accordingly the Fourier transform of the payoff function
has large high-frequency terms. The Carr-Madan solution is in fact
to dampen the weight of the high frequencies by multiplying the
payoff by an exponential decay function. This will lower the
importance of the singularity, but at the cost of degradation of the
solution accuracy.

As the Eqs.~(\ref{HEuropPrice}) can be used whenever the
characteristic function of the given model is known, we can apply it
to the Black-Scholes model as well that gives us the Black-Scholes
option price $V^{BS}$ which is a well known analytic expression. Now
the idea is to rewrite representation of the option price in
the Eqs.~(\ref{HEuropPrice}) in the form

\begin{equation}\label{newRepres}
V^M = [V^M - V^{BS}] + V^{BS}.
\end{equation}

The term in braces can now be computed with FFT as
\begin{eqnarray}\label{NewEuropPrice}
\Pi_1^{M-BS} &=& \dfrac{\xi}{2\pi}
\int_{-\infty}^{\infty}\dfrac{e^{-iu \kappa} \left[ \phi_{X_t} (u-i) e^{i(u-i)\omega T}  -
\phi_{BS} (u-i)e^{- \frac{\sigma ^2}{2}T}\right] }{i u} du, \\
\Pi_2^{M-BS} &=& \dfrac{\xi}{2\pi} \int_{-\infty}^{\infty}
\dfrac{e^{- iu \kappa } \left[\phi_{X_t} (u)e^{iu \omega T} - \phi_{BS} (u)\right]}{i u}du, \nn \\
V^{M} - V^{BS} &=& \xi \left[e^{-q T}S_0 \Pi_1^{M -BS} - e^{-r T} K
\Pi_2^{M - BS}\right], \nn
\end{eqnarray}

\ni where $\kappa = \ln (K/S) - (r-q)T$, $\phi_{BS} (u) = \exp\left(-\frac{\sigma ^2 T}{2}u^2 \right)$
and $\phi _{X_T}(-i) = e^{-\omega T}$. This is possible
because we have removed the divergence in the integrals. In addition
the magnitude of $\phi_{X_t} (z) - \phi_{BS} (z)$ is smaller than
that of $\phi_{X_t} (z)$ that increases accuracy of the solution.

In more detail, first terms of the expansion of $\phi_{X_t} (u)e^{iu \omega T} - \phi_{BS} (u)$ and
$\phi_{X_t} (u-i) e^{i(u-i)\omega T}  - \phi_{BS} (u-i)e^{- \frac{\sigma ^2}{2}T}$ in series at small $u$ are
\begin{eqnarray} \label{expan1}
D_1|_{u=0} &\equiv& \phi_{X_t} (u)e^{iu \omega T} - \phi_{BS} (u) = T ( \theta  + \omega + \dfrac{\sigma^2}{2} )i u + O(u^2) \\
D_2|_{u=0} &\equiv&\phi_{X_t} (u-i) e^{i(u-i)\omega T}  - \phi_{BS} (u-i)e^{- \frac{\sigma ^2}{2}T} =
- \left( \sigma ^2 + \frac{\theta +\sigma ^2}{-1 + \nu(\theta + \sigma ^2/2) } - \omega \right)iu + O(u^2) \nn
\end{eqnarray}

However, an usage of these expressions in the  Eq.~(\ref{NewEuropPrice}) together with the FFT method produces
an error of the order of $O(u)$. That is why it is better to choose a small $u=\epsilon $, for instance $\epsilon =10^{-6}$,
then computing integrands in the Eq.~(\ref{NewEuropPrice}) exactly and substituting
$D_{1,2}|_{u=0} \approx D_{1,2}|_{u=\epsilon }$.

\begin{figure}[bht]
\begin{flushleft}
\begin{minipage}[ht]{0.4\textwidth}
\includegraphics[totalheight=2in]{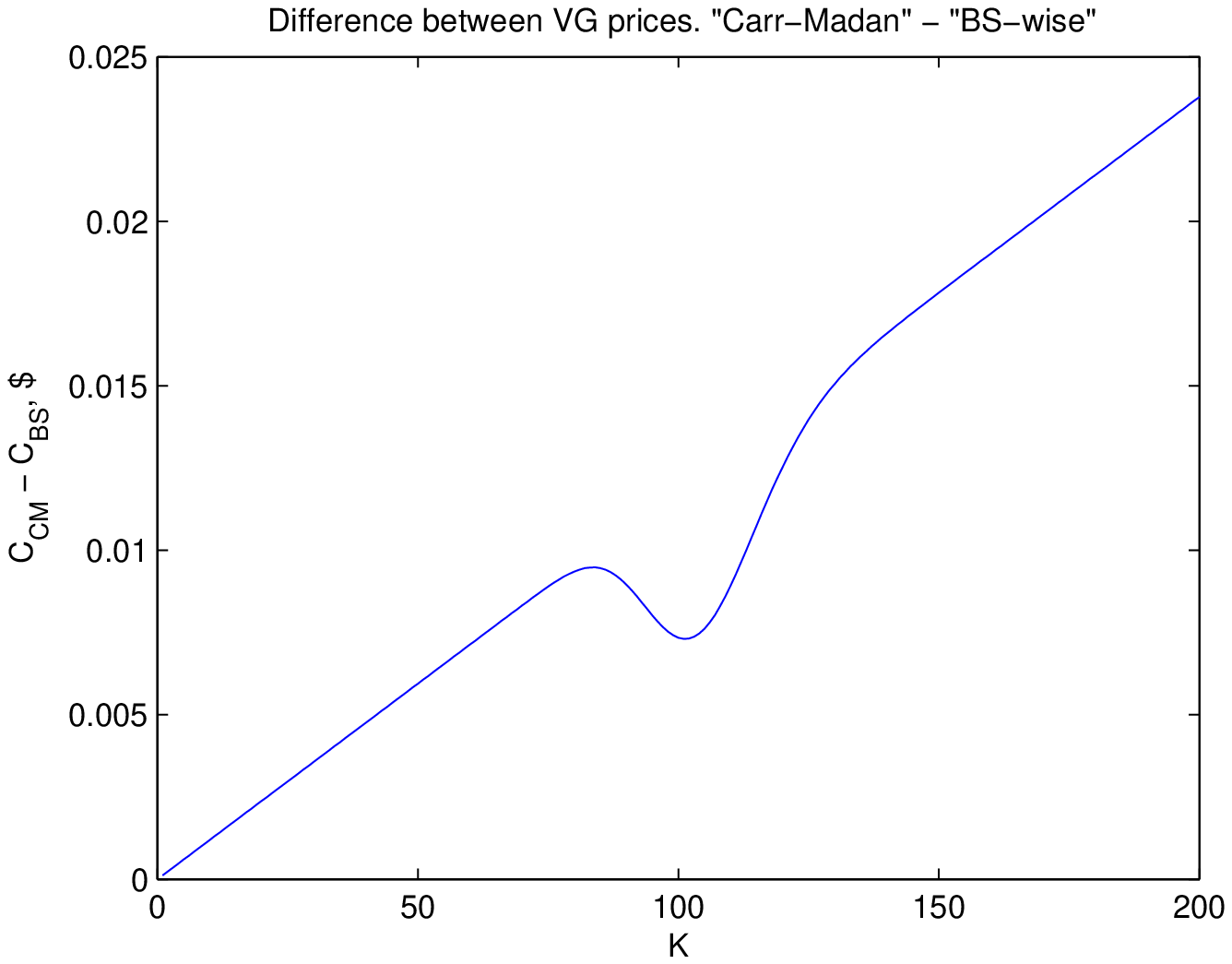}
\caption{European option values in VG model. Difference between the Carr-Madan solution and Black-Scholes-wise solution
with $D_{1,2}(u=0)$ at $T=1.0 yr, \sigma = 0.1, \theta =0.1, \nu = 0.1$}
\label{BSFFT1}
\end{minipage}
\hspace{0.1\textwidth}
\begin{minipage}[ht]{0.4\textwidth}
\includegraphics[totalheight=2in]{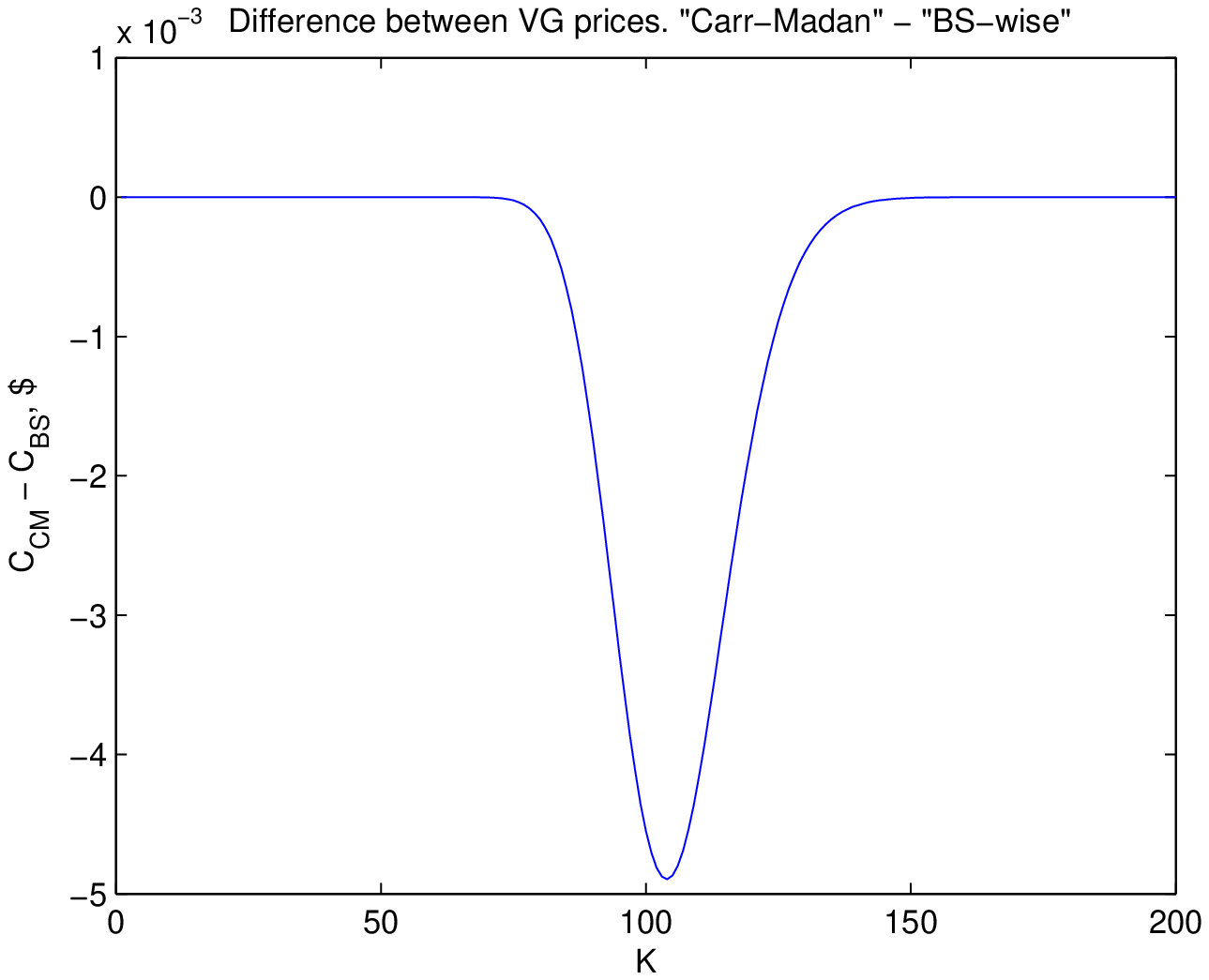}
\caption{European option values in VG model. Difference between the Carr-Madan solution and Black-Scholes-wise solution
with $D_{1,2}(u=\epsilon)$ at $T=1.0 yr, \sigma = 0.1, \theta =0.1, \nu = 0.1$}
\label{BSFFT2}
\end{minipage}
\end{flushleft}
\end{figure}

Fig.~\ref{BSFFT1}, \ref{BSFFT2} show the results of our computation of the European option values
under the VG model. Difference between the Carr-Madan solution and Black-Scholes-wise solution
with $D_{1,2}(u=\epsilon)$ and $D_{1,2}(u=0)$ at $T=1.0 yr, \sigma = 0.1, \theta =0.1, \nu = 0.1$ are plotted for 200 strikes.
It is seen that for the first method the difference is of the order of 0.5\%.

\section{Convergency and performance}

Artur Sepp reported in \cite{Sepp2003} that the convergency of the
Black-Scholes-wise method is approximately 3 times faster than that
of the Lewis method. It could be understood because as we mentioned
above in the limit of small $u$ the difference between the VG
solution and the Black-Scholes formula which is under the Fourier
integral is of the second order in $u$ while in the Lewis method it
is of the zero order. In other words using the Black-Scholes-wise
formula allows us to remove a part of the FFT error instead substituting it
with the exact analytical solution of the Black-Scholes problem.

\begin{figure}[bht]
\begin{flushleft}
\begin{minipage}[ht]{0.4\textwidth}
\includegraphics[totalheight=2in]{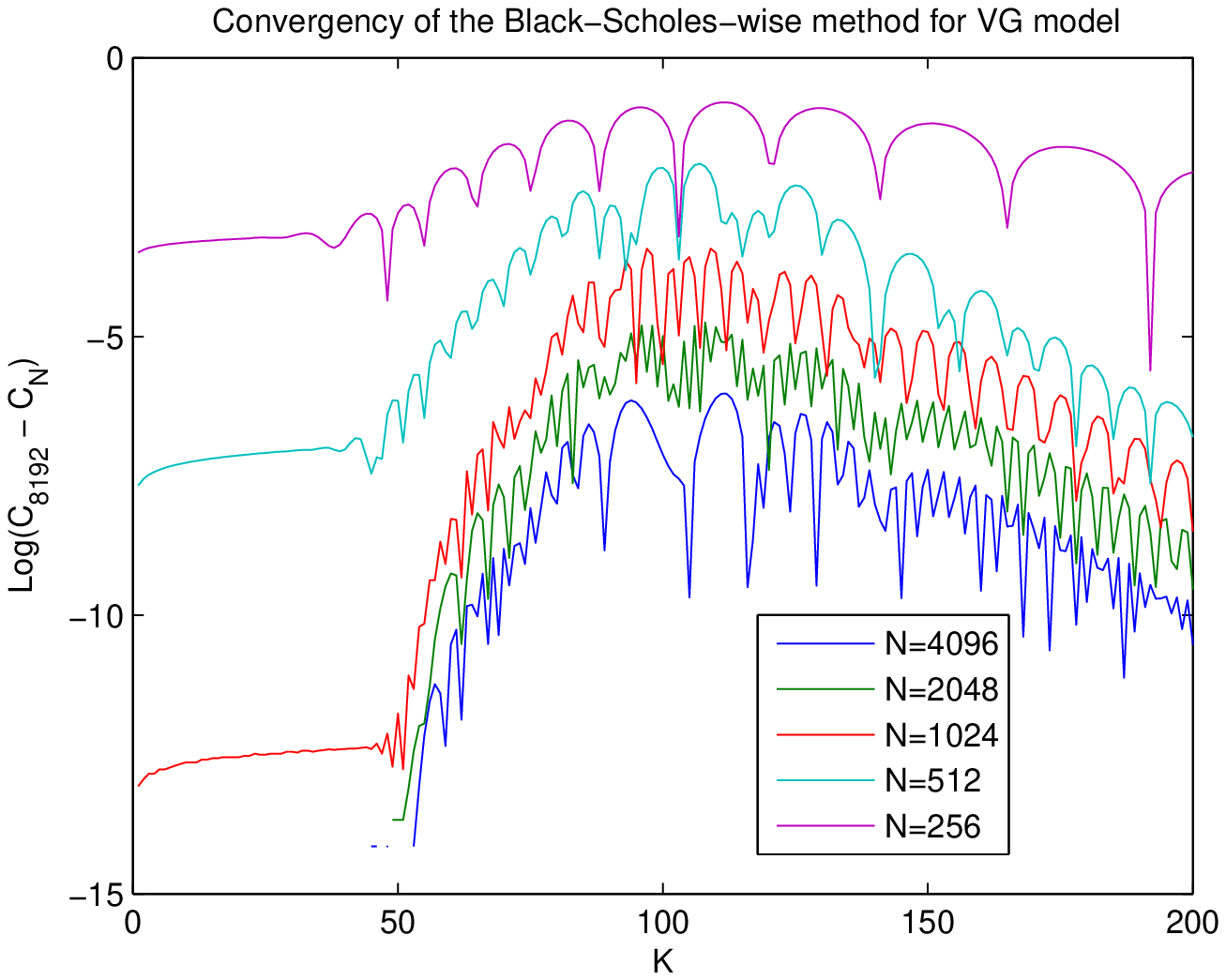}
\caption{Convergency of the Black-Scholes-wise method. Difference between the
option price obtained with $N=8192$, and that with $N=4096, 1024,512,256$}.
\label{convBS}
\end{minipage}
\hspace{0.1\textwidth}
\begin{minipage}[ht]{0.4\textwidth}
\includegraphics[totalheight=2in]{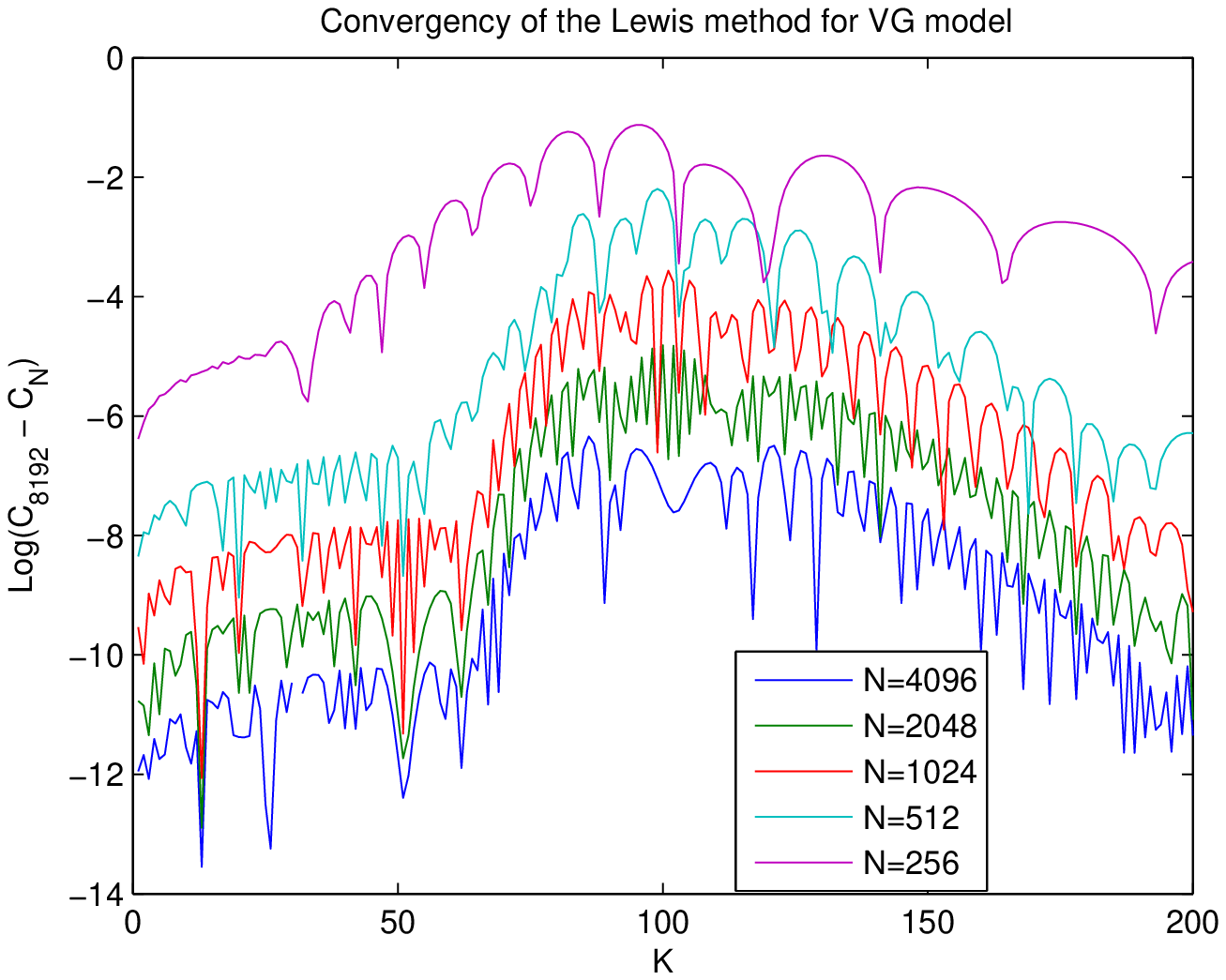}
\caption{Convergency of the Lewis method. Difference between the
option price obtained with $N=8192$, and that with $N=4096, 1024,512,256$}.
\label{convLewis}
\end{minipage}
\end{flushleft}
\end{figure}

\begin{figure}[bht]
\begin{flushleft}
\begin{minipage}[ht]{0.4\textwidth}
\includegraphics[totalheight=2in]{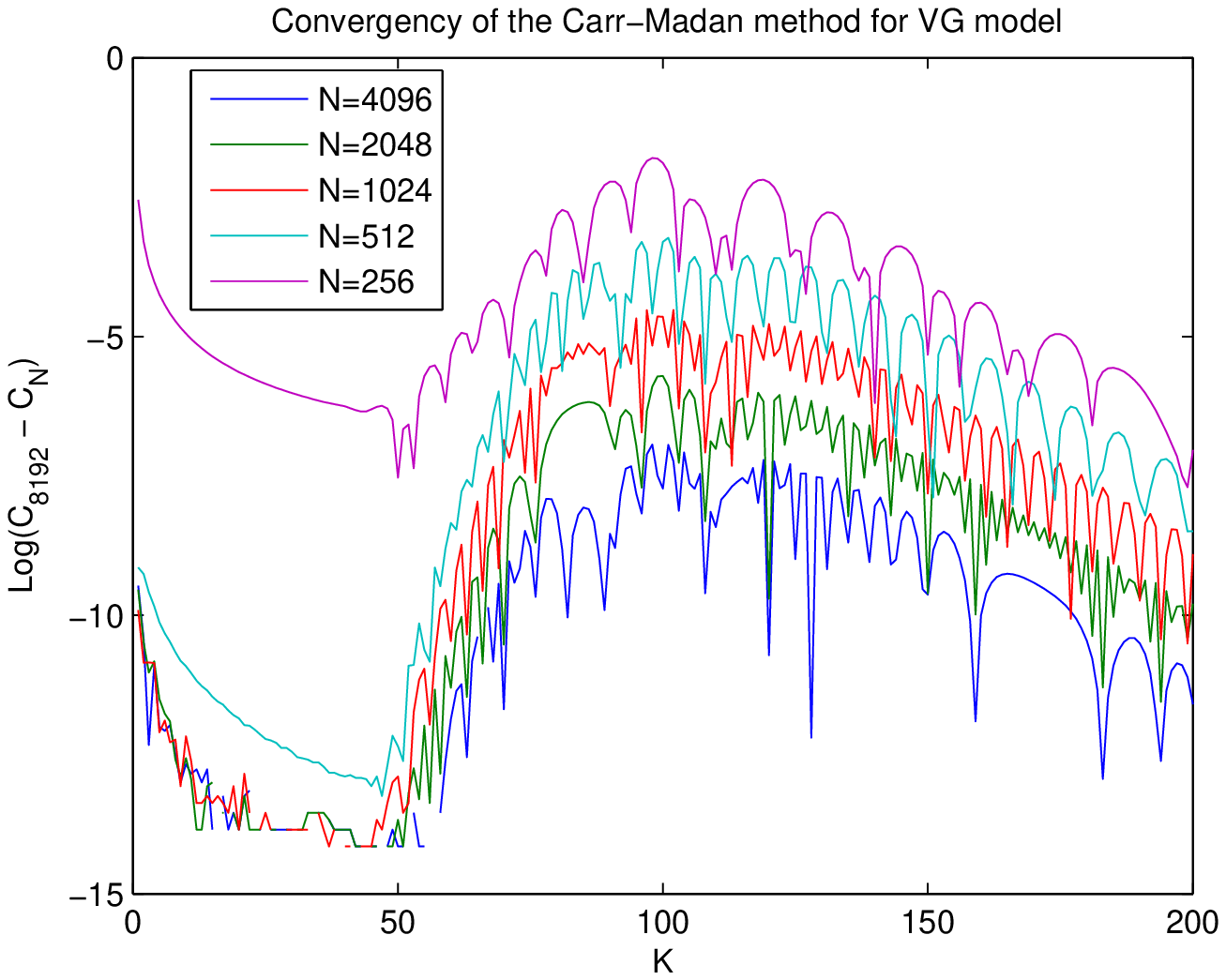}
\caption{Convergency of the Carr-Madan method. Difference between the
option price obtained with $N=8192$, and that with $N=4096, 1024,512,256$}.
\label{convCM}
\end{minipage}
\hspace{0.1\textwidth}
\begin{minipage}[ht]{0.4\textwidth}
\includegraphics[totalheight=2in]{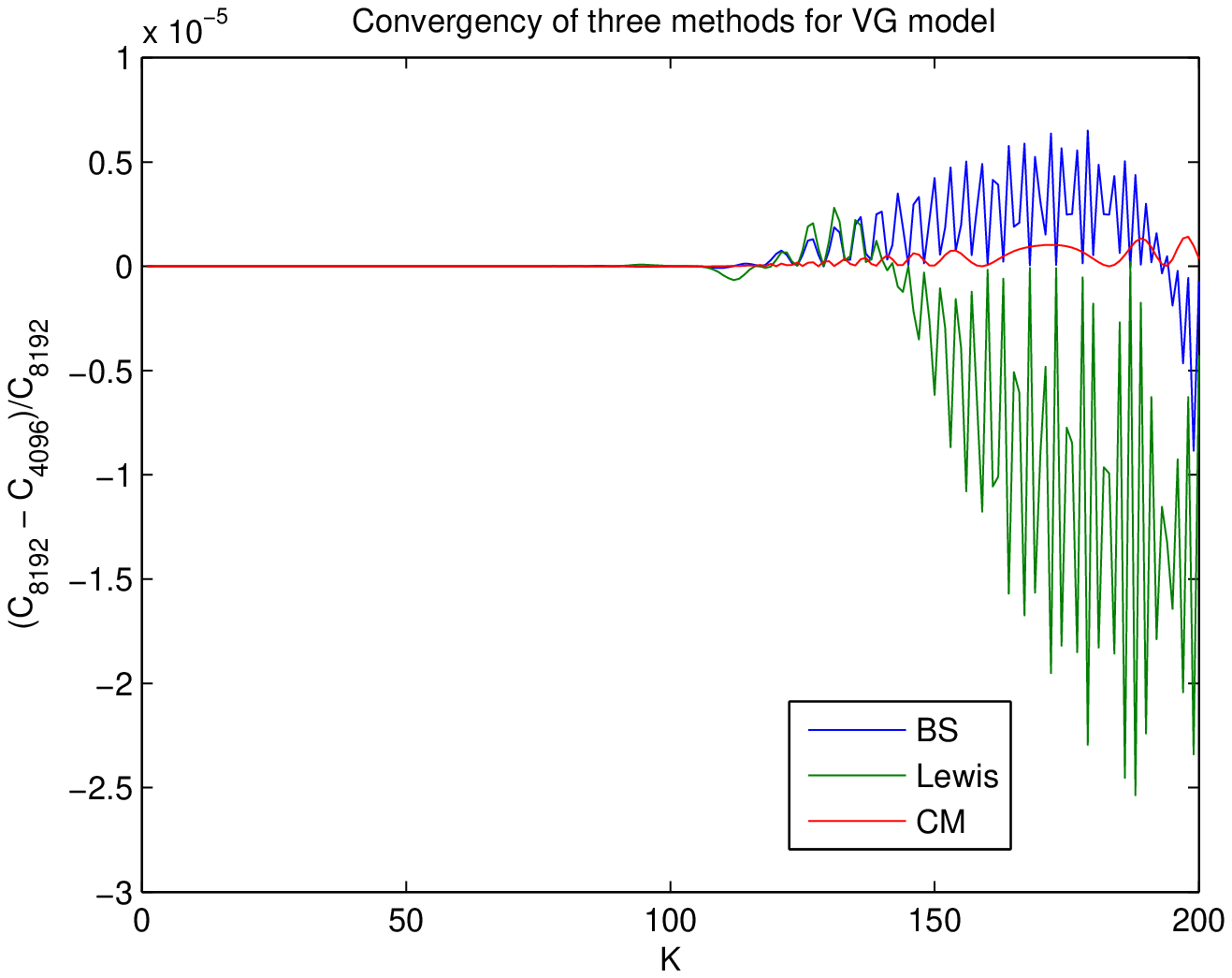}
\caption{Convergency of all three methods}.
\label{conv3}
\end{minipage}
\end{flushleft}
\end{figure}

We also fulfilled investigation of how all three methods converge for the VG model.
The results are given in Fig.~\ref{convBS},\ref{convLewis},\ref{convCM}. We display $\log_{10}$
difference between the option price obtained with $N=8192$, and that with $N=4096, 1024,512,256$.
We don't see much difference in the convergency of the Lewis and Black-Scholes-wise method while
the Carr-Madan methods behaves better at low $N$. In Fig.~\ref{conv3} we also present the ratio
$(C_{N=8192} - C_{N=4096})/C_{N=8192}$ for all three methods. The Carr-Madan still converges better for
out of the money spot prices while convergency of two other methods is similar.

Cont and Tankov also analyze the Lewis method. They emphasize the
fact that the integral in the Eq.~(\ref{altFFT}) is much easier to
approximate at infinity than that in the Carr-Madan method, because
the integrand decays exponentially (due to the presence of
characteristic function). However, the price to pay for this is
having to choose $\mu _1$. This choice is a delicate issue because
choosing big $\mu _1$ leads to slower decay rates at infinity and
bigger truncation errors and when $\mu _1$ is close to one, the
denominator diverges and the discretization error becomes large. For
models with exponentially decaying tails of Levy measure, $\mu _1$
cannot be chosen a priori and must be adjusted depending on the
model parameters.

Carr and Madan in \cite{CarrMadan:99a} compare performance of 3 methods for computing VG prices:
VGP which is the analytic formula in Madan, Carr, and Chang;
VGPS which computes delta and the risk-neutral probability of finishing
in-the-money by Fourier inversion of the distribution function, i.e. according to the
Eq.~(\ref{HEuropPrice}); VGFFTC which is a Carr-Madan method using FFT to invert the dampened call price;
VGFFTTV which uses FFT to invert the modified time value. The results are given in Tab.~(\ref{comparison}). The computation times for the first
two methods involve 160 strike levels. The first 4 rows of Tab.~(\ref{comparison}) display
4 combinations of parameter settings, while the last 4 rows show computation times in seconds.

\begin{table}[ht]
\begin{flushleft}
\begin{minipage}[ht]{0.4\textwidth}
\begin{tabular}{|l|r|r|r|r|}
\hline
& case 1 & case 2 & case 3 & case 4 \\
\hline
$\sigma$ & .12 & .25 & .12 & .25 \\
$\nu$ & .16 & 2.0 & .16 & 2.0 \\
$\theta$ & -.33 & -.10 & -.33 & -.10 \\
$T$ & 1 & 1 & .25 & .25 \\
\hline
VGP & 22.41 & 24.81 & 23.82 & 24.74 \\
VGPS & 288.50 & 191.06 & 181.62 & 197.97 \\
VGFFTC & 6.09 & 6.48 & 6.72 & 6.52 \\
VGFFTTV & 11.53 & 11.48 & 11.57 & 11.56 \\
\hline
\end{tabular}
\caption{CPU times for VG pricing. Represented from  \cite{CarrMadan:99a}. }
\label{comparison}
\end{minipage}
\hspace{0.1\textwidth}
\begin{minipage}[ht]{0.4\textwidth}
\begin{tabular}{|l|r|r|r|r|}
\hline
& case 1 & case 2 & case 3 & case 4 \\
\hline
$\sigma$ & .12 & .25 & .12 & .25 \\
$\nu$ & .16 & 2.0 & .16 & 2.0 \\
$\theta$ & -.33 & -.10 & -.33 & -.10 \\
$T$ & 1 & 1 & .25 & .25 \\
\hline
Lewis & 0.031 & 0.031 & 0.031 & 0.031 \\
Carr-Madan & 0.047 & 0.047 & 0.032 & 0.032 \\
BS-wise & 0.078 & 0.078 & 0.062 & 0.062 \\
\hline
\end{tabular}
\caption{CPU times for VG pricing. Our calculations.}
\label{OurCalc}
\end{minipage}
\end{flushleft}
\end{table}

It is seen that the analytic formula is slow while the slowest (and least
accurate in case 4) method inverts for the delta and for the probability of
paying off.

However, this is not true if one uses a modified method given in the Eq.~(\ref{NewEuropPrice}).
Our calculations show that the performance of the Lewis method is same as the Carr-Madan method, and
the performance of the Black-Scholes-wise method is only twice worse (because we need 2 FFT to compute 2 integrals)
(see Tab.~\ref{OurCalc}).

\section{Conclusion}

We discussed various analytic and numerical methods that have been
used to get option prices within a framework of VG model. We showed
that a popular Carr-Madan's FFT method \cite{CarrMadan:99a} blows up
for certain values of the model parameters even for European vanilla option. Alternative methods -
one originally proposed by Lewis, and Black-Scholes-wise method were
considered that seem to work fine for any value of the VG
parameters. Convergency and accuracy of these methods is comparable with that of the Carr-Madan
method, thus making them suitable for being used to price options with the VG model.

\newpage
\newcommand{\noopsort}[1]{} \newcommand{\printfirst}[2]{#1}
  \newcommand{\singleletter}[1]{#1} \newcommand{\switchargs}[2]{#2#1}

\end{document}